# Role of Oxygen Vacancies in Stabilizing the Orthorhombic Phases of $Hf_{0.5}Zr_{0.5}O_2$ Nanoparticles

Yuriy O. Zagorodniy[1], Eugene A. Eliseev[1], Valentin V. Laguta[1,2], Petr Jiricek[2], Jana Houdkova[2], Lesya D. Demchenko[3,4], Oksana V. Leshchenko[1], Victor N. Pavlikov[1], Lesya P. Yurchenko[1], Anna O. Diachenko[5], Michail D. Volnyanskii[5], Oleksandr S. Pylypchuk[6], Myroslav V. Karpets[4], Mikhail P. Trubitsyn[2,5*], Dean R. Evans[7†], and Anna N. Morozovska[6‡]

[1] Frantsevich Institute for Problems in Materials Science of the National Academy of Sciences of Ukraine, 3, str. Omeliana Pritsaka, 03142 Kyiv, Ukraine

[2] Institute of Physics of the Czech Academy of Sciences, Na Slovance 1999/2, 18200 Prague 8, Czech Republic

[3] Stockholm University, Department of Chemistry, Sweden

[4] Ye. O. Paton Institute of Materials Science and Welding, National Technical University of Ukraine "Igor Sikorsky Kyiv Polytechnic Institute", 37, Beresteisky Avenue, Kyiv, Ukraine, 03056

[5] Oles Honchar Dnipro National University, 72, Nauky Avenue, 49000 Dnipro, Ukraine

[6] Institute of Physics of the National Academy of Sciences of Ukraine, 46, Nauky Avenue, 03028 Kyiv, Ukraine

[7] Zone 5 Technologies, Special Projects, San Luis Obispo CA 93401, USA

## Abstract

In this work we study the stabilization of the orthorhombic phases in small $Hf_{0.5}Zr_{0.5}O_2$ nanoparticles (average size ~ 7 nm) annealed under different oxygen partial pressures. Concentration of the oxygen vacancies, which is determined by annealing conditions, was estimated from the electron paramagnetic resonance spectra and X-ray photoelectron spectroscopy. The fraction of the orthorhombic phases, that is determined by the X-ray diffraction and nuclear magnetic resonance, depends on the concentration of oxygen vacancies. Phenomenological calculations based on Landau-Ginzburg-Devonshire theory considering trilinear coupling between nonpolar, antipolar, and polar phonon modes, indicate that chemical strain induced by oxygen vacancies can stabilize the orthorhombic o-III phase with the ferroelectric long-range ordering in small $Hf_{0.5}Zr_{0.5}O_2$ nanoparticles. The theory confirms the stability of ferroelectric polarization in the vacancy-enriched $Hf_{0.5}Zr_{0.5}O_2$ nanoparticles. The increase in the intensity of the dielectric permittivity maximum observed near 350 – 380 K in a PVDF matrix embedded with $Hf_{0.5}Zr_{0.5}O_2$ nanoparticles annealed in the $CO+CO_2$ atmosphere is clearly associated with an increase in oxygen vacancy concentration. The vacancies lead to defect-induced elastic dipole formation and an increase in ionic conductivity, which decreases the depolarization field and may induce the ferroelectric-like phase transition in the vacancy-enriched $Hf_{0.5}Zr_{0.5}O_2$ nanoparticles. Due to the interfacial effects, the negative capacitance states may be realized in weakly screened and spatially isolated $Hf_{0.5}Zr_{0.5}O_2$ nanoparticles embedded in the PVDF matrix. The present approach based on oxygen-vacancy-induced elastic and screening effects may provide a route for engineering ferroelectric-like states in other nanoscale ferroic oxides.

* Corresponding author, e-mail: trubitsyn_m@ua.fm

† Corresponding author, e-mail: deanevans@zone5tech.com

‡Corresponding author, e-mail anna.n.morozovska@gmail.com

## I. INTRODUCTION

The discovery of ferroelectricity in thin films of hafnia-zirconia ($Hf_xZr_{1-x}O_2$, $0 \leq x \leq 1$) [1, 2, 3] has made these materials highly promising candidates for non-volatile memory devices and integrated layered capacitors [4, 5, 6]. The chemical stability and full compatibility with complementary metal-oxide-semiconductor (CMOS) processes make $Hf_xZr_{1-x}O_2$ thin films particularly attractive for modern microelectronic applications. $Hf_xZr_{1-x}O_2$ thin films are successfully used as gate dielectrics in field-effect transistors because they do not exhibit many of the limitations inherent to perovskite ferroelectrics, which suffer from poor compatibility with silicon technologies due to substantial lattice mismatches and the presence of interfacial dead layers that degrade their ferroelectric response. However, $Hf_xZr_{1-x}O_2$ thin films are among the most enigmatic ferroic materials, since the physical mechanisms responsible for their ferroelectric, ferrielectric, and/or antiferroelectric behaviors, which emerge with the film thickness decrease, remain under active debate [7, 8].

In contrast to thin films, bulk $Hf_xZr_{1-x}O_2$ (as well as $HfO_2$ and $ZrO_2$) is a nonpolar centrosymmetric material. At room temperature and normal atmospheric pressure ($\sim 10^5$ Pa), these structurally equivalent compounds stabilize in the monoclinic phase (m-phase) with the space group $P2_1/c$. With increasing temperature, this phase undergoes a transformation to the tetragonal phase (t-phase) with the space group $P4_2/nmc$, followed by a transition to the cubic phase (c-phase) with the space group *Fm3m* [9, 10]. Separately, under high pressure, two orthorhombic phases (nonpolar o-I and antipolar o-II) with the space groups *Pbca* and *Pnma* can be stable [11, 12]. Note that bulk $Hf_xZr_{1-x}O_2$ materials stabilized in the o-I or o-II phases do not undergo a transition to the ferroelectric phase under doping, temperature, additional pressure, or other external factors.

The ferroelectric (FE) properties of $Hf_xZr_{1-x}O_2$ thin films are associated with the formation of a polar orthorhombic o-III phase with the non-centrosymmetric space group $Pca2_1$. Formation of the FE o-III phase was observed for the first time in 9-nm $Hf_xZr_{1-x}O_2$ ($0 \leq x \leq 1$) films covered by TiN electrodes [1]. In this experiment, undoped $HfO_2$ films had monoclinic symmetry and paraelectric properties, $ZrO_2$ films had tetragonal symmetry and antiferroelectric properties, whereas $Hf_xZr_{1-x}O_2$ ($0 < x < 1$) thin films demonstrated distinct ferroelectric properties, with maximal remanent polarization observed in $Hf_{0.5}Zr_{0.5}O_2$ films. The formation and stabilization of the FE o-III phase in $Hf_xZr_{1-x}O_2$ thin films can be attributed to several factors, including elastic strain arising from deposition on substrates with different lattice parameters [13, 14, 15], changes in the electronic structure due to doping with Si [16], Y [17] or Al [18] ions, lattice distortions caused by the formation of oxygen vacancies [19, 20], size effects [21], and the interplay of these different mechanisms.

The important role of oxygen vacancies in the stabilization of the FE o-III phase in thin $HfO_2$ films has been confirmed by numerous experimental [22, 23] and theoretical [24] studies. In particular, it was demonstrated experimentally [22] that the ferroelectric properties of undoped 15 nm thick $HfO_2$ films deposited on Pt, TiN, and Si substrates and annealed in a nitrogen atmosphere were determined by the concentration of oxygen vacancies, as follows from the X-ray photoelectron spectroscopy analysis. It was later shown experimentally that oxygen-deficient $Hf_xZr_{1-x}O_{2-y}$ nanoparticles, annealed under a $CO+CO_2$ atmosphere, contain a large fraction (> 70 %) of orthorhombic phases (o-I, o-II, and likely o-III), which were indistinguishable by X-ray diffraction because of the broadening of the diffraction peaks associated with the small sizes (<10 nm) of the nanoparticles. The presence of the FE o-III phase in the $Hf_xZr_{1-x}O_{2-y}$ nanoparticles is very likely because they exhibit ferroelectric-like properties [25, 26], such as a colossal dielectric response over a wide frequency range [27], resistive switching, and pronounced charge accumulation [28].

Furthermore, it was shown theoretically [24] that the influence of vacancies on the stabilization of the FE o-III phase in $HfO_2$ films arises from the formation of elastic dipoles around point defects, which leads to the local breaking of inversion symmetry. Density functional theory (DFT) calculations performed in Ref. [23] confirm the influence of oxygen vacancies on the phase stability of $HfO_2$ films and revealed that the FE o-phase becomes energetically more favorable than the m-phase, although it remains less favorable than the tetragonal and cubic phases. However, the physical mechanisms governing the appearance and stabilization of the o-phases (including nonpolar o-I, antipolar o-II, and polar o-III phases) in $Hf_xZr_{1-x}O_2$ nanoparticles are very poorly studied.

In this work, we experimentally study the stabilization of the o-phases in $Hf_xZr_{1-x}O_2$ nanoparticles with different concentrations of oxygen vacancies. The choice of nanoparticles is motivated by the desire to eliminate the influence of the substrate, thus isolating size effects and defect-related effects. The concentration of oxygen vacancies was estimated by electron paramagnetic resonance (EPR) and X-ray photoelectron spectroscopy (XPS), whereas the phase composition was estimated by X-ray diffraction (XRD) and nuclear magnetic resonance (NMR). Phenomenological calculations based on Landau-Ginzburg-Devonshire (LGD) theory considering trilinear coupling between nonpolar, antipolar, and polar phonon modes in $Hf_xZr_{1-x}O_2$ nanoparticles were performed to assess the stabilization of the FE o-III phase by the chemical strain induced by oxygen vacancies. The obtained results indicate that $Hf_xZr_{1-x}O_2$ nanoparticles can serve as a model system for disentangling the intrinsic roles of oxygen vacancies and size effects in stabilizing ferroelectric orthorhombic phases relevant to thin films and nanoparticle-based composites.

## II. EXPERIMENTAL RESULTS AND DISCUSSION

### A. Samples Preparation and Experimental Techniques

To study how oxygen vacancies affect the emergence of the o-phases (including the desirable FE o-III phase), we prepared $Hf_{0.5}Zr_{0.5}O_2$ nanoparticles by solid-state organonitrate synthesis. The process employed aqueous mixtures of zirconium and hafnium nitrate salts ($ZrO(NO_3)_2{\cdot}2H_2O$ and $Hf(NO_3)_2{\cdot}2H_2O$) dissolved in distilled water at concentrations not exceeding 5 – 10 % (for methodological details, see Refs. [25, 26]). The $Hf_{0.5}Zr_{0.5}O_2$ nanoparticles labeled as sample N1 were annealed in air at 700 °C for 6 hours. Sample N2 consists of $Hf_{0.5}Zr_{0.5}O_2$ nanoparticles annealed at 500°C for 6 hours in a $CO + CO_2$ atmosphere to strongly increase the concentration of oxygen vacancies.

Composite materials prepared from $Hf_{0.5}Zr_{0.5}O_2$ nanoparticles obtained from samples N1 and N2 were embedded in the PVDF polymer matrix. The samples were prepared from pre-compressed (100 MPa) composites in the form of irregularly shaped plates with linear dimensions of 5 – 8 mm and a thickness of about 0.5 – 0.8 mm. The $Hf_{0.5}Zr_{0.5}O_2$ nanopowder (taken from sample N1 or N2) occupied approximately 13 vol. % of the total composite volume. Silver electrodes were deposited on the faces of the samples. The electrical properties of the $Hf_{0.5}Zr_{0.5}O_2$ – PVDF composites were measured in an AC field ($f \approx 10^2 - 10^6$ Hz) using a Keysight E4980AL LCR meter. Measurements were performed from room temperature up to ~440 K, above which the PVDF-based composites significantly softened due to heating, making reliable measurements impossible.

X-ray photoelectron spectra (XPS) were measured by an AXIS Supra photoelectron spectrometer using monochromatic Al Kα radiation (1486.6 eV, 300 W, area analyzed – 0.7×0.3 mm$^2$). A source of gas cluster ions was used to remove contaminants from the surface. To minimize damage of the sample surface due to interaction with high-energy ions, the ion source operated in the cluster mode (the energy of the primary cluster ion was 5 keV, and the number of argon atoms in the cluster was 2000). This mode of the ion source significantly reduces surface damage of the samples. The irradiation time using primary cluster ions was 5 minutes. Photoelectron spectra were recorded before and after the etching process. The main peak of the C 1s spectrum, corresponding to carbon contamination (C–C), was used as a reference to calibrate the binding energy scale (i.e., fixed at 284.5 eV).

NMR spectra were collected at room temperature using a Bruker (Avance II) 9.4 T commercial spectrometer. Extremely broad $^{91}$Zr spectra were acquired by stepping the spectrometer frequency across the spectrum and recording spin echoes using the conventional $90_x - \tau - 90_y - \tau$ spin echo pulse sequence.

The spin-echo delay time $\tau$ was 20 $\mu$s and the repetition delay time between scans was 0.1 s. A four-phase "exorcycle'' phase sequence (xx, xy, x–x, x–y) was used to form echoes with minimal distortions due to antiechoes, ill-refocused signals, and piezoresonances [29]. The length of the $\pi/2$ pulse was $t_{\pi/2} = 2$ $\mu$s. Individual echoes were Fourier transformed and then superimposed in the frequency domain. Spectral simulations were performed using the TopSpin and MRSimulator software programs [30].

EPR spectra were recorded using a Bruker Elexsys E580 spectrometer operating in the X-band (9.8 GHz) with a magnetic field modulation of 100 kHz. The effective g-values were determined according to the relation $g_{eff} = h\nu/(\mu_B H_r)$, where $h$ is Planck's constant, $\nu$ is the microwave frequency, $\mu_B$ is the Bohr magneton, and $H_r$ is the resonance field. Spectral simulations and fitting parameters were obtained using with the EasySpin software package [31].

## B. Materials Characterization

According to the scanning electron microscopy (SEM) analysis (see **Fig. S1** in Supplementary Materials [32]), samples N1 and N2 are homogenous and consist of particle clusters with an average size of about 65 nm. The electron dispersion spectra (EDS) analysis confirmed that the constituent elements (Hf, Zr, and O) are uniformly distributed in the $Hf_{0.5}Zr_{0.5}O_2$ nanoparticles, forming a solid solution (see **Fig. S2** in Supplementary Materials). At the same time, the appearance of Debye-Scherrer rings and transmission electron microscopy (TEM) images show that the $Hf_{0.5}Zr_{0.5}O_2$ nanoparticles have individual sizes around 7 nm and a high degree of crystallinity (see **Fig. 1**).

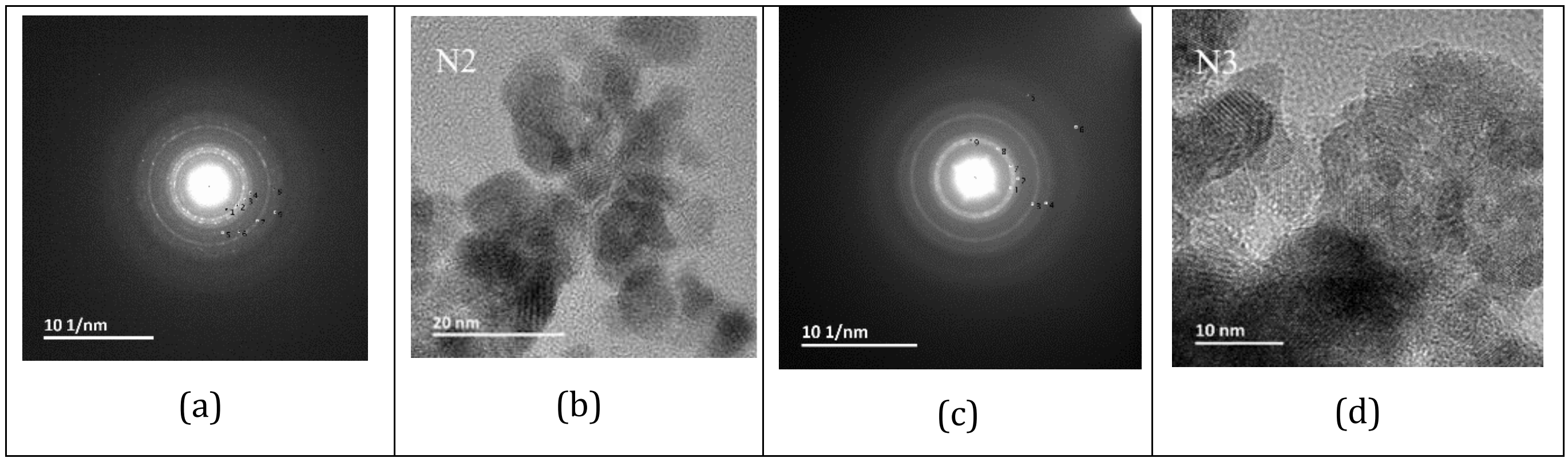

**Figure 1.** TEM images of $Hf_{0.5}Zr_{0.5}O_2$ nanoparticles annealed in an air atmosphere, sample N1 **(a, b)** and in $CO + CO_2$ atmosphere, sample N2 **(c, d)**.

X-ray diffraction (XRD) measurements of the samples are presented in **Fig. S3** (see Supplementary Materials). Despite the high crystallinity of the nanoparticles, their XRD spectra exhibit broad diffraction lines arising from significant surface effects, thereby complicating the analysis. While the spectrum of the monoclinic phase differs significantly from other possible phases, the spectra of the tetragonal and orthorhombic phases are very difficult to distinguish from each other using XRD. Three orthorhombic phases, which can be nonpolar o-I, antipolar o-II, and polar FE o-III phases (according to symmetry theory analysis), are even more difficult to distinguish, as they have almost identical cation arrangements and differ only in the positions of the oxygen atoms [33].

Based on the full-profile analysis of the XRD spectra, the $Hf_{0.5}Zr_{0.5}O_2$ nanopowders annealed in air (sample N1) predominantly display a coexistence of the m-phase (63.54 wt. %) and the o-phases (36.46 wt. %). Sample N2, consisting of $Hf_{0.5}Zr_{0.5}O_2$ nanopowders annealed in the $CO+CO_2$ atmosphere, reveals only o-phases (100 wt. %). However, according to Ref. [33], in the case of sample N1, the residual error between the "monoclinic + orthorhombic" basis and the "monoclinic + tetragonal + orthorhombic" basis used for assignment of the XRD peaks is usually small. To achieve a clearer distinction between the tetragonal and orthorhombic phases, nuclear magnetic resonance (NMR) measurements were performed.

Distinguishing among the three orthorhombic phases, namely the nonpolar o-I, antipolar o-II phases, and polar FE o-III phases, is particularly challenging. However, one can assume that the mechanism of formation of the orthorhombic phases under the influence of oxygen vacancies promotes the formation of the polar o-III phase. The assumption is corroborated by the following arguments. The unit cell of the nonpolar o-I phase can be imagined as consisting of two coupled polar cells distinguished by oxygen displacements in opposite directions [34]. The polar o-III phase arises from the displacement of one type of oxygen atoms in the same direction along the z-axis. Displacement in the opposite direction can also induce a FE phase with opposite polarity. The antipolar o-II phase arises from the displacement of the same type of oxygen atoms in different directions along the z-axis. Since oxygen vacancies produce elastic strains, which stabilize the shift of oxygen atoms in the same direction along the z-axis [24], it is reasonable to assume that the vacancies stabilize the FE o-III phase.

### C. X-ray Photoelectron Spectroscopy Results

To study variations in the chemical composition and valence states of surface species depending on the synthesis conditions, the samples were examined using X-ray photoelectron spectroscopy **(**XPS). To distinguish between possible surface defects and contamination, the Hf 4f, Zr 3d, and O 1s XPS spectra were obtained for both annealed samples (N1 and N2) before after surface etching with argon ions for 300 seconds.

The Hf 4f and Zr 3d XPS spectra consist of two main peaks corresponding to the spin-orbit splitting of the 4f and 3d electrons, respectively. Fitting each component of the Hf 4f doublet by Gaussian peaks yields binding energies of 18.2 and 16.5 eV (±0.1 eV) for Hf $4f_{5/2}$ and Hf $4f_{7/2}$, respectively, which correspond to the Hf-O bond, where Hf is in the +4 oxidation state. Fitting the Zr 3d doublet gives 182.1 and 184.5 eV (±0.1 eV) for $3d_{5/2}$ and $3d_{3/2}$, respectively, corresponding to $Zr^{4+}$ in $ZrO_2$. The positions of these peaks are virtually identical for all samples. Any visible shoulder with a lower binding energy that could correspond to a metal (Hf or Zr) in the +3 oxidation state is absent in the XPS spectra.

The main difference in the spectra is the shoulder with a higher binding energy, whose intensity depends on sample preparation conditions and decreases substantially after etching the sample surface. **Figure 2** shows the Hf 4f and Zr 3d XPS spectra of the sample N2, $Hf_{0.5}Zr_{0.5}O_2$ annealed in the $CO+CO_2$ atmosphere, before and after surface etching. In both the Hf and Zr spectra, the high binding energy shoulder is substantially more pronounced before etching and decreases after etching, indicating a strong surface-related contribution. Such high-energy shoulders in transition metal spectra (i.e., Hf and Zr) are usually associated with shake-up satellites [35, 36]. Their intensity depends on the degree of covalency of the chemical bond formed by the metal atoms with the functional groups present at the surface, which may be influenced by oxygen vacancy modified local environments; it may also be related to surface contamination during the synthesis process. Partial removal of surface contaminants and weakly bound groups significantly reduces the intensity of this peak. Correlation of the intensity of the shake-up satellites with the synthesis conditions likely arises from the enhanced ability of vacancy sites to adsorb oxygen species in samples annealed in a $CO + CO_2$ atmosphere. A more detailed analysis presented below (see **Fig. 3**) suggests that the high-energy shoulder is linked, at least indirectly, to reducing conditions (oxygen vacancy formation), because the O 1s signal likely reflects surface-related oxygen species or oxygen vacancy-modified local environments rather than the oxygen vacancies themselves.

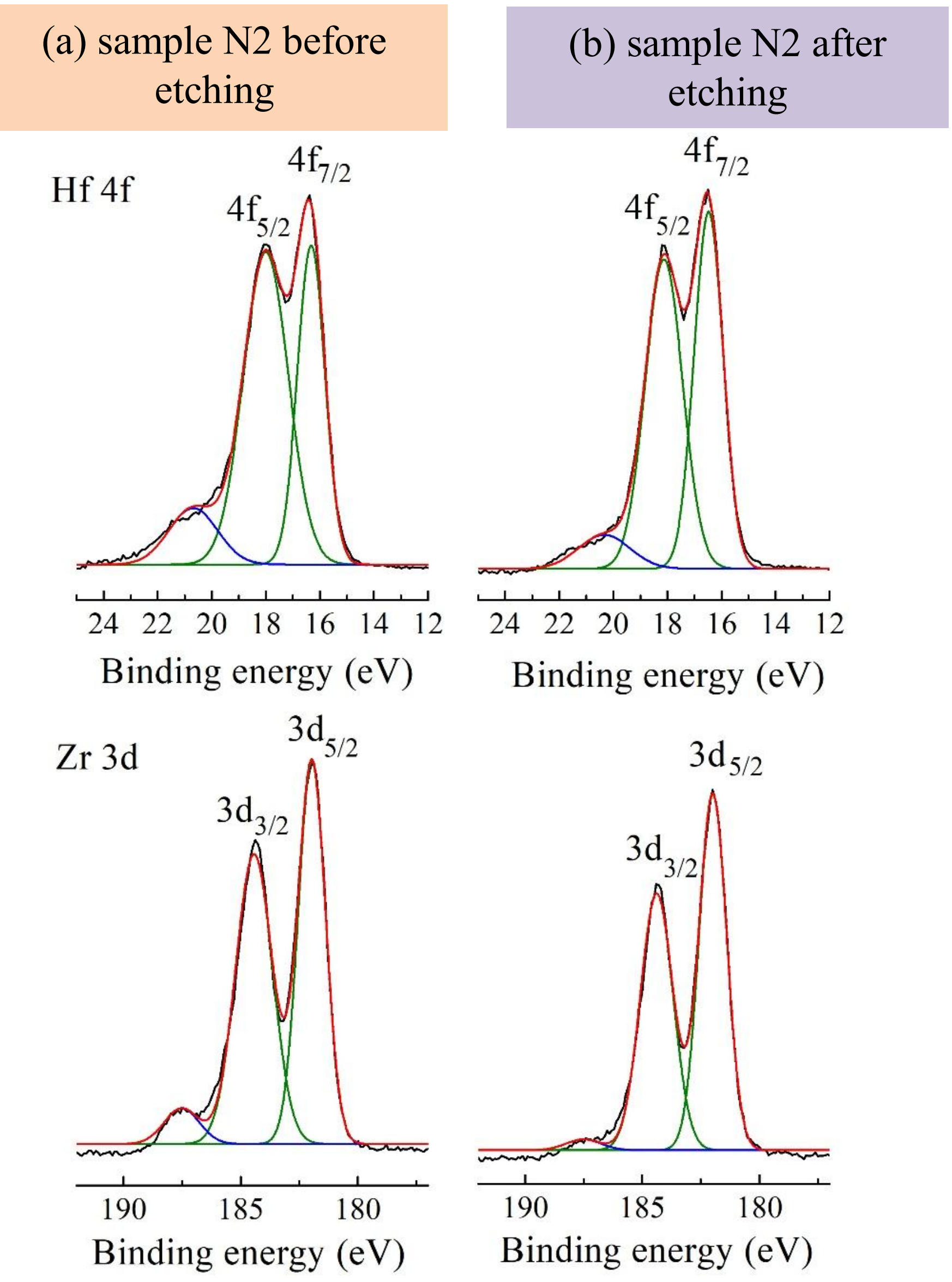

**Figure 2.** Hf 4f and Zr 3d XPS spectra of $Hf_{0.5}Zr_{0.5}O_2$ nanoparticles annealed in a $CO+CO_2$ atmosphere (sample N2). Column **(a)** corresponds to sample N2 before etching, and column **(b)** corresponds to sample N2 after 300 s of surface etching. Black curves represent experimentally measured spectra, and red curves are fits using three Gaussian components. The two green components correspond to the spin-orbit doublet, while the blue component represents the higher binding energy shoulder.

To obtain more details about the formation of chemical bonds by oxygen atoms and, possibly, to probe the formation of oxygen vacancies, a detailed study of the high-resolution O 1s XPS spectra was performed (see **Fig. 3**). The O 1s XPS spectra of the annealed samples (**Fig. 3**, column **(a)**) show a highly asymmetric peak, which can be decomposed into three Gaussian components with binding energies of about 530, 531.5, and 532.2 – 533.8 eV. They are designated as I, II, and III in **Fig. 3**. As can be seen from the figure, the shape of the O 1s peaks depends on the synthesis conditions. The binding energies of O 1s electrons for all components obtained from the Gaussian decomposition, as well as their relative intensities, are presented in **Table 1**, which also provides the contribution of the orthorhombic phases in the studied samples, determined from the XRD spectra. The change in the number of defects presumably induced by synthesis conditions is accompanied by a change in the asymmetry of the spectral peak, manifested in a redistribution of intensity between peaks II and III, with an increase in peak III and a corresponding decrease in peak II. This effect is particularly pronounced for peak III.

From **Figure 3**, the high-binding-energy peak III is more pronounced in the samples annealed in $CO+CO_2$ than in the samples annealed in air, accompanied by a corresponding redistribution of spectral intensity between peaks II and III. The effect is particularly noticeable after etching. This behavior suggests that peak III is linked, at least indirectly, to reducing conditions and oxygen vacancy-related local environments. The $CO + CO_2$ atmosphere likely acts as a reducing environment by lowering the oxygen partial pressure and thereby promoting oxygen vacancy formation, whereas annealing in air has an opposite effect and suppresses oxygen vacancy formation (see **Table 1**).

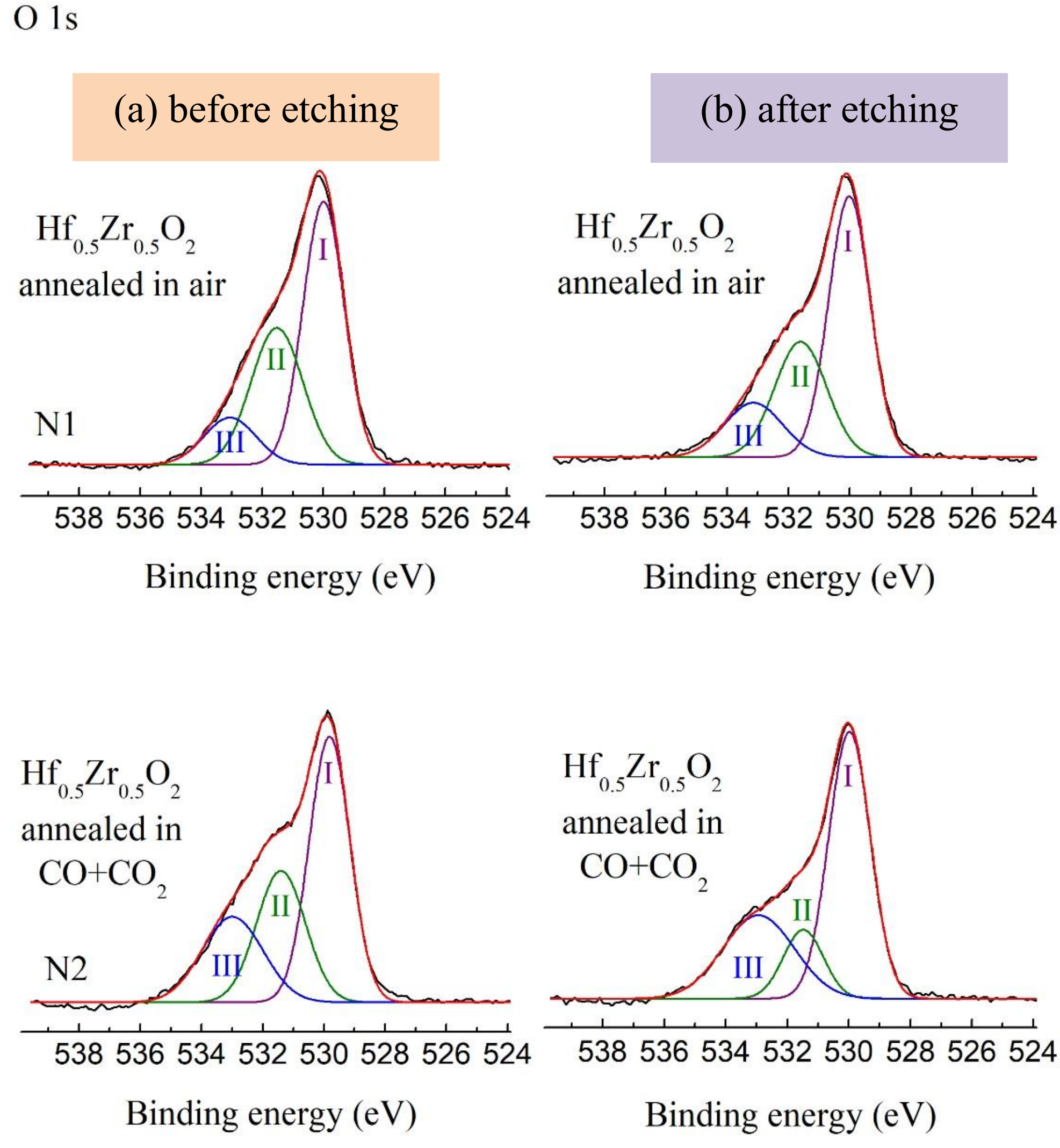

**Figure 3.** O 1s XPS spectra of the annealed $Hf_{0.5}Zr_{0.5}O_2$ samples, N1 and N2, before etching **(a)** and after surface etching for 300 s **(b)**. Black curves are experimentally measured spectra, red curves are fits using three Gaussians components shown by the violet, green, and blue curves.

**Table 1.** Preparation conditions of the $Hf_{0.5}Zr_{0.5}O_2$ nanopowders, binding energies, peak areas obtained from Gaussian decomposition of the O1s XPS spectra, and the fraction of the o-phase for each sample. The XPS peaks discussed in detail in the text are attributed to: I - lattice oxygen; II - defect-related surface oxygen; III - high binding energy peak commonly associated with surface species such as hydroxyl groups and vacancy-related defects.

| Nanopowder samples | | Binding Energy (eV) | | | Peak Area (%) | | | Fraction of o-phases o-I, o-II and o-III (wt. %, determined from XRD) |
|---|---|---|---|---|---|---|---|---|
| | | I | II | III | I | II | III | |
| N1 $Hf_{0.5}Zr_{0.5}O_2$ | Annealed in air | 530 | 531.5 | 533.1 | 53 | 35 | 12 | 36 |
| | Etched | 530 | 531.8 | 533.1 | 54 | 30 | 16 | |
| N2 $Hf_{0.5}Zr_{0.5}O_2$ | Annealed in $CO+CO_2$ | 529.8 | 531.4 | 533 | 48 | 29 | 23 | 100 |
| | Etched | 530 | 531.5 | 532.9 | 56 | 14 | 30 | |

The most intense in these compounds corresponds to the peak with a binding energy of about 530 eV and can be assigned to lattice oxygen (peak I), which has similar values in the $HfO_2$ and $ZrO_2$ [37, 38], where the Hf and Zr ions have an oxidation state of +4. Peaks II and III, by their position, can be attributed to hydroxyl groups or various carbon groups, such as physically adsorbed $CO_2$ molecules, carbonates formed as a result of the interactions between CO and oxygen at the surface, as well as defects of the C-O-*Me* type, where the metal *Me* refers to hafnium or zirconium. At the same time, there are many studies in which the O 1s peaks with higher binding energy are considered to be characteristic of oxygen vacancies present in the samples [39, 40,41, 42, 43, 44]. In this case, peaks II and III refer to O 1s signals from oxygen-deficient regions and surface oxygen vacancies capturing oxygen-related groups (e.g., OH groups) [43, 44]. Several explanations have been proposed for the sensitivity of O 1s spectra to oxygen vacancies. For example, the formation of trivalent metal ions ($Hf^{3+}$ or $Zr^{3+}$) can reduce the electron density on neighboring oxygen atoms, leading to an increase in their O 1s binding energy [39]. This may contribute to the observed increase in the high-binding-energy component (peak III), although other surface-related species may also play a role.

While this approach is widely accepted in literature, the attribution of this O 1s peak to the oxygen deficient regions has also attracted serious criticism. One of the arguments is that oxygen molecules, water molecules and/or oxygen-related groups adsorbed from the surrounding atmosphere immediately oxidize the hafnia-zirconia surface with oxygen vacancies presented in their structure. This leads to the healing of the surface oxygen vacancy and results in the appearance of O 1s signals in the region of 531.5 – 533 eV, which is attributed to absorbed water and hydroxyl groups formed in result of dissociation of adsorbed water [45]. Also, the DFT calculations for different metal oxides show that oxygen vacancies have a very small effect on the O 1s binding energies of neighboring oxygen atoms, while OH groups can shift the core levels of the O 1s peak by more than 2 eV [46]. In the present interpretation, peak I is associated with lattice oxygen, peak II with surface-related and/or defect-related oxygen environments, while peak III appears more strongly connected to reducing conditions and oxygen-vacancy-modified environments.

Despite the controversy surrounding the assignment of oxygen vacancies from the O1s peak and the lack of a clear physical basis for the proposed explanations, there is a clear correlation between the shape of the XPS line and the oxygen vacancy concentrations expected for samples prepared under different annealing conditions. The presence of vacancies, as well as $Hf^{3+}$ and $Zr^{3+}$ ions, is not definitively identified in the Hf 4f and Zr 3d XPS spectra but may be more reliably confirmed by EPR studies (see the next section). EPR is a bulk-sensitive volumetric technique, whereas XPS probes nanoparticles located within the sample surface layer (~5-10 nm), where the chemical environment is strongly influenced by ambient exposure. EPR provides information on vacancy-related paramagnetic defects (e.g., oxygen vacancies, $Hf^{3+}$ and $Zr^{3+}$ centers), while XPS reflects the chemical states of oxygen at the surface. Taken together, the XPS and EPR data provide a more complete picture by linking the electronic signature of defects (EPR) with the corresponding chemical response at the surface (XPS).

The relative contribution of peaks I–III changes noticeably in the XPS spectra of the etched samples (**Fig. 3**, column **b**). For both annealed samples, a sharp decrease in the intensity of peak II is observed (see **Table 1**), which allows us to conclude that this peak includes signals from surface contaminants as well as from weakly bound molecules on the sample surface. The effect is particularly pronounced for the $Hf_{0.5}Zr_{0.5}O_2$ sample annealed in the CO + $CO_2$ atmosphere. Meanwhile, the relative intensity of peak III with the higher binding energy increases after etching and reaches a maximum for the $Hf_{0.5}Zr_{0.5}O_2$ sample annealed in the CO+$CO_2$ atmosphere. This result can be explained by the role of oxygen vacancies, which act as active surface centers and promote the adsorption of water molecules. Some of these molecules subsequently dissociate into OH groups, depending on the number of electrons (from 0 to 2) captured by the vacancy. This also implies that water and OH groups adsorbed by the sites of oxygen vacancies are bound to the surface more strongly. It can be used as a tool for a rough quantitative estimate of the number of vacancies. However, it is impossible to obtain an accurate estimation, since water is adsorbed not only by the sites of oxygen vacancies but also at other sites on the surface.

The relative intensity of peak III for the $Hf_{0.5}Zr_{0.5}O_2$ sample annealed in the $CO+CO_2$ atmosphere is approximately twice as high as that for all other samples. Assuming that the concentration of oxygen vacancies scales approximately linearly with the peak intensity, one may conclude that annealing in $CO+CO_2$ atmosphere increases the vacancy concentration significantly. At the same time, this sample exhibits 100% o-phase content, as determined from the XRD data. This correlation supports the proposed assignment and is consistent with the role of oxygen vacancies in stabilizing o-phases in the studied samples.

We attempted to estimate the increase in the vacancy concentration at the surface of the sample annealed in a $CO+CO_2$ atmosphere, which contains 100 % of o-phases, compared to the sample annealed in air. This was done by assuming that OH groups and water molecules adsorbed at the vacancy sites on the surface, producing a separate peak (located at a different binding energy than the main lattice oxygen peak). An increase in its intensity directly indicates an increase in the number of vacancies. Using this way, we estimated that the $Hf_{0.5}Zr_{0.5}O_2$ sample N2 annealed in the $CO+CO_2$ atmosphere may contain approximately 10 – 15 % oxygen vacancies.

### D. Electron Paramagnetic Resonance Spectroscopy

Electron paramagnetic resonance **(**EPR) is a powerful technique for characterizing defects that have an unpaired spin. Such defects include oxygen vacancies with a single trapped electron. Moreover, the formation of oxygen vacancies is accompanied by a reduction in the valence state of neighboring Hf or Zr cations. This reduction can lead to the presence of uncompensated 5d or 4d electrons on Hf and Zr, giving rise to EPR signals corresponding to $Hf^{3+}$ and $Zr^{3+}$ ions. These spectral features are well documented in the literature [47, 48] and can be effectively used to estimate the concentration of oxygen vacancies in the investigated samples.

EPR spectra of $Hf_{0.5}Zr_{0.5}O_2$ nanoparticles without purposeful oxygen vacancy formation have been reported in Ref. [28]. In the present work, the spectra of the hafnia-zirconia compounds show significant differences primarily in samples containing varying concentrations of oxygen vacancies and other synthesis-induced defects. **Figure 4** presents the spectra of $Hf_{0.5}Zr_{0.5}O_2$ nanoparticles annealed in air (sample N1) and in a $CO+CO_2$ atmosphere (sample N2).

The effective g factors were calculated using the equation, $g_{ef} = h\nu/\mu_B H_r$, where $h$ is Planck's constant, $\nu$ is the operating frequency, $\mu_B$ is the Bohr magneton, and $H_r$ is the magnetic resonant field. The intensities of the lines presented in **Fig. 4** are not normalized to the number of spins; instead they are chosen for the convenience of presenting the spectra in a single figure.

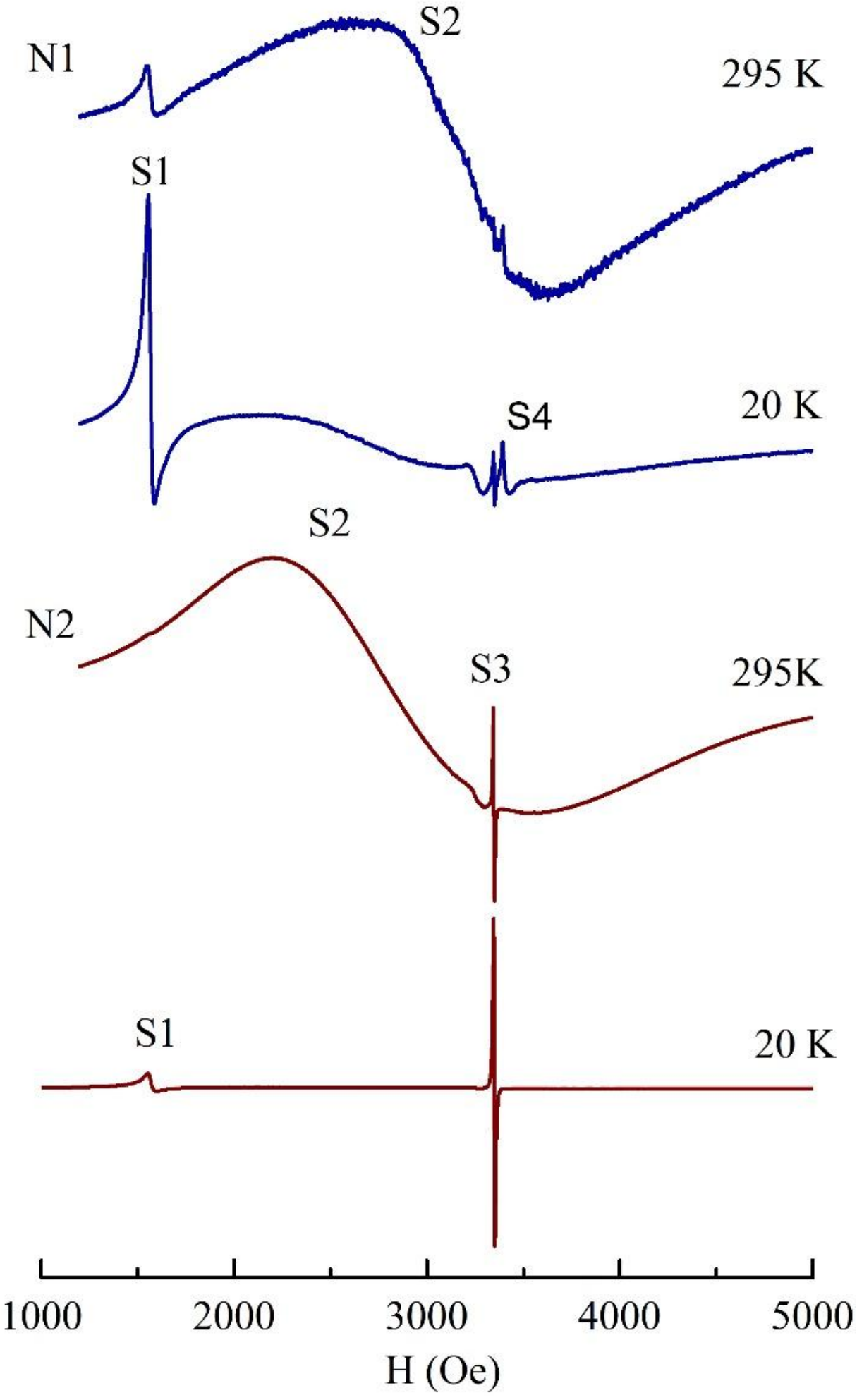

**Figure 4.** EPR spectra of $Hf_{0.5}Zr_{0.5}O_2$ nanopowders annealed in air (sample N1) and $CO+CO_2$ (sample N2) atmospheres, each recorded at temperatures 20 K and 295 K.

The EPR spectra of the studied compounds generally exhibit four lines of varying intensity, depending on the sample and the measurement temperature. Line S1, with a g-factor of approximately 4.3, corresponds to isolated $Fe^{3+}$ ions in a rhombically distorted environment [49] and is attributed to unintentional $Fe^{3+}$ impurities present in all samples. The broad line S2, with a g-factor in the range of 2.0 – 2.4, is attributed to paramagnetic ions with uncompensated electron spins. These ions form clusters where dipole–dipole and exchange interactions between individual spins contribute to the observed line broadening. The presence of this line in both types of samples, together with the expected high density of oxygen vacancies in the surface layer of nanoparticles annealed in air or $CO+CO_2$, suggests that S2 is primarily associated with clusters formed in the surface region.

The sharp line S3, with a g-factor of 2.0031, appears in all compounds but shows markedly higher intensity in the sample annealed in $CO+CO_2$. This line can be attributed to oxygen vacancies containing a trapped electron [47], as well as carbon radicals remaining after synthesis. To exclude contributions from organic radicals, whose EPR signals typically saturate at low temperatures, spectra were recorded while gradually lowering the sample temperature. It was observed that the relative intensity of S3 increases significantly upon cooling. **Figure 5** illustrates the spectra at low temperature (20 K). Both samples exhibit partial saturation of the S2 line, whereas the S3 line remains unsaturated and dominates the spectrum of sample N2, which, based on its synthesis conditions, is expected to contain a large number of oxygen vacancies. The absence of saturation, together with a narrow linewidth (peak-to-peak width of ~5 G), indicates that exchange interactions between paramagnetic centers are sufficiently strong to produce exchange narrowing, thereby confirming the presence of a high concentration of oxygen vacancies in this compound.

At the same time, in the spectrum of sample N1, where the intense S3 line does not dominate, an additional feature S4 is clearly resolved at low temperature. Detailed EPR spectra of both samples, obtained in a narrow range around the S4 feature, are presented in **Fig. 5**. The spectrum of sample N1 recorded at 20 K, in addition to the moderate-intensity S3 line, reveals two additional distinct features: a line with $g_x = 1.999$, $g_y = 2.008$, and $g_z = 2.004$, which can be attributed to hole-type oxygen centers [50], and a line with $g_x = 1.948$, $g_y = 1.975$, and $g_z = 1.967$. The latter line, whose theoretical simulation is shown at the bottom of the figure, corresponds to isolated $Hf^{3+}$ and $Zr^{3+}$ ions [28, 48, 50].

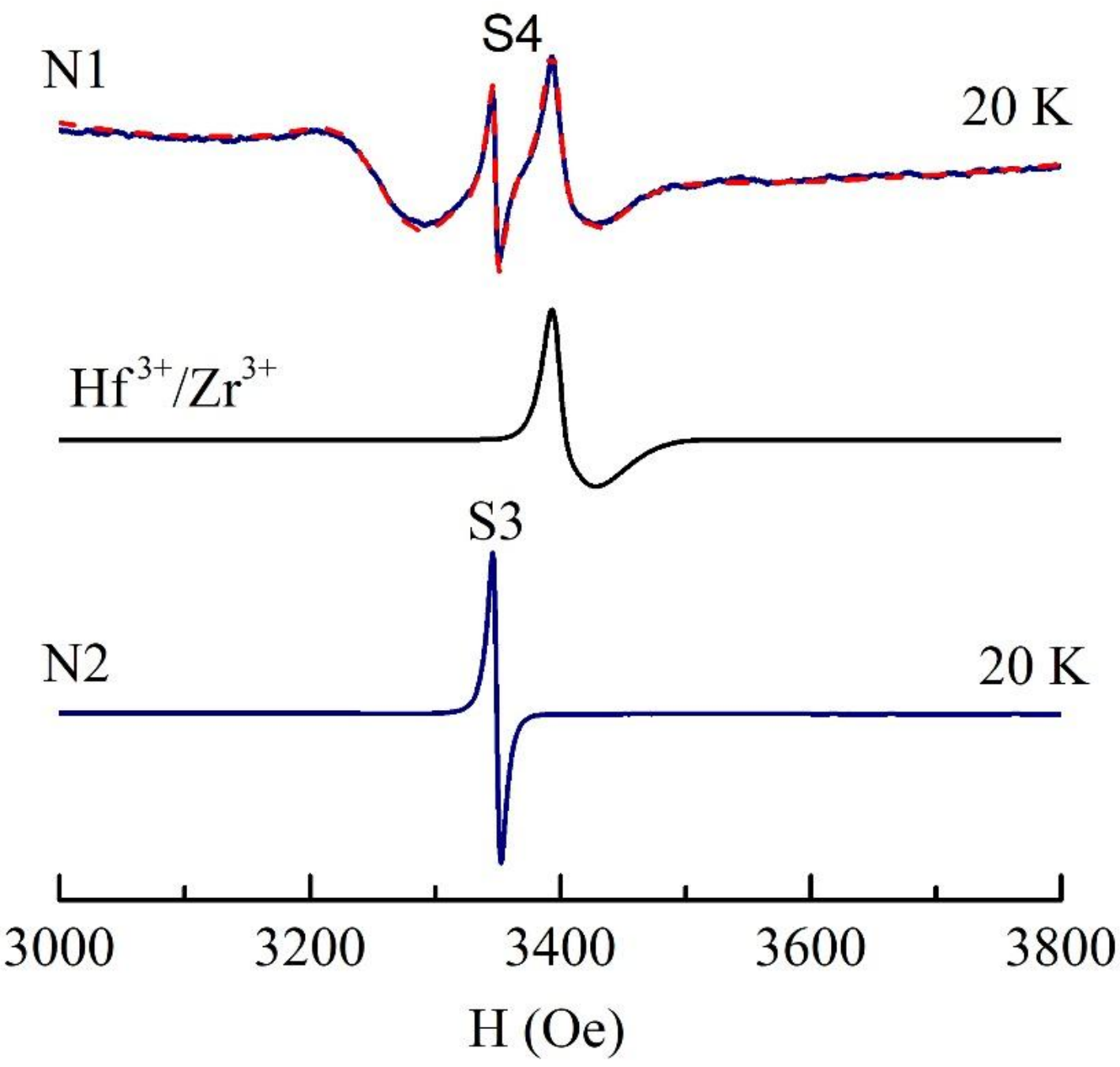

**Figure 5.** Detailed EPR spectra of $Hf_{0.5}Zr_{0.5}O_2$ nanoparticles annealed in air (sample N1) and $Hf_{0.5}Zr_{0.5}O_2$ nanoparticles annealed in $CO+CO_2$ (sample N2), obtained at 20 K, together with the fitting of the spectrum of sample N1 (red dashed line) and the fitting component corresponding to $Hf^{3+}$ and $Zr^{3+}$ ions.

Thus, sample N2, which contains a large concentration of oxygen vacancies, intentionally introduced through annealing, demonstrates the formation of clusters in which the density of paramagnetic ions, such as oxygen vacancies with trapped electrons and $Hf^{3+}/Zr^{3+}$ ions, is sufficient to produce an exchange-narrowed line. This signal dominates the EPR spectra across all temperatures, down to 20 K.

### E. Nuclear Magnetic Resonance Spectra of $^{91}$Zr

Nuclear magnetic resonance (NMR) spectra provide a powerful method to study the symmetry of the local atomic environment; it is widely used to determine the phase composition of materials [51, 52]. The most suitable isotope for NMR studies of $Hf_{0.5}Zr_{0.5}O_2$ nanoparticles is the $^{91}$Zr, which has a natural abundance of 11.2 %, a nuclear spin $I$ = 5/2, and a quadrupole moment $eQ$ = -17.6 fm$^2$.

As demonstrated in studies of zirconia-based materials [52], the shape of the $^{91}$Zr NMR spectra in zirconium oxides is primarily determined by the quadrupolar interaction of the zirconium nuclei with the electric field gradient (EFG) generated by the surrounding charges. The spin Hamiltonian of the quadrupolar interactions can be expressed as [53]:

$$\boldsymbol{H}_Q = \frac{C_Q h}{4I(2I-1)}\left[3\boldsymbol{I}_z^2 - I^2 + \frac{\eta}{2}(\boldsymbol{I}_+ + \boldsymbol{I}_-)\right], \quad (1)$$

where $C_Q = eQV_{zz}/h$ is the quadrupole coupling constant, proportional to the largest eigenvalue $|V_{zz}|$ of the EFG tensor in its principal axes system with $|V_{zz}| > |V_{xx}| > |V_{yy}|$, and $\eta = (V_{xx} - V_{yy})/V_{zz}$ is the asymmetry parameter of the EFG tensor.

Only the central ½ ↔ -½ transition, which is affected by a second-order perturbation of the Zeeman interaction, is well resolved in the spectra of studied samples. The frequency shift of the central transition due to the quadrupole interaction can be expressed as:

$$\nu_{1/2}^{(2)} = -\frac{\nu_Q^2}{16\nu_L}\left(I(I+1) - \frac{3}{4}\right) f_\eta(\theta, \varphi), \quad (2)$$

where $\nu_Q = -\frac{3C_Q}{2I(2I-1)}$, and the function $f_\eta(\theta, \varphi)$ depends on the asymmetry of the EFG tensor and the orientation of its principal axes relative to the external magnetic field [54].

The central transition line in the $^{91}$Zr NMR spectra typically exhibits a width on the order of several hundred kilohertz; this is true even for bulk samples [52, 55]. NMR lines from higher-order satellite transitions in powdered samples are broadened due to the much stronger first-order quadrupolar interaction and the distribution of quadrupolar parameters arising from local structural distortions. Nevertheless, the overall shape of the $^{91}$Zr NMR spectra reflects the symmetry of a crystalline phase, which determines the local symmetry of the electric field gradient (EFG) at the zirconium nuclei. For example, in a phase with tetragonal symmetry, the zirconium site exhibits η = 0 and a quadrupole coupling constant $C_Q$ = 19.1 MHz, whereas in the monoclinic phase η increases slightly to 0.1 [52]. In contrast, specially prepared magnesia-partially stabilized zirconia (MgO-PSZ), containing 45.5 % of the o-phase as confirmed by neutron diffraction, exhibits noticeably different parameters: $C_Q$ = 17 MHz and η = 0.8, where the large asymmetry parameter η significantly changes the spectral line shape [52].

The experimental $^{91}Zr$ NMR spectrum of $Hf_{0.5}Zr_{0.5}O_2$ nanoparticles, annealed in a $CO+CO_2$ atmosphere, is presented in **Fig. 6a**. The best fit of the spectrum requires two components, reflecting the presence of distinct regions with different distributions of quadrupolar parameters around the zirconium nuclei. Regions in close proximity to bulk and surface defects exhibit a broad distribution of quadrupolar parameters, producing a wide line in the spectrum that is well approximated by a Gaussian function (see **Fig. 6(a)**, dashed curve). In contrast, regions with a more moderate distribution of quadrupolar parameters exhibit an orthorhombic local symmetry. Their NMR response can be modeled using Eq.(1) with $C_Q$= 17.8 MHz and η = 0.8 (see **Fig. 6(b)**). These values are in good agreement with those previously reported for the o-phase of MgO-PSZ [52].

To emphasize the distinction between the orthorhombic and tetragonal phases, **Fig. 6(c)** shows a line constructed using the same quadrupole coupling constant $C_Q$ and other parameters as in **Fig. 6(b)**, but with η = 0, corresponding to the EFG of zirconium in a tetragonal environment. The pronounced difference in line shapes supports the correct assignment of the experimental spectrum to the o-phases.

It should be noted that a broad NMR line with a Gaussian-type background (see **Fig. 6(a)**, dashed curve) is typically observed in disordered materials such as glasses, but it may also result from a broad distribution of local environments in crystalline systems. To describe such distributions of the EFG in disordered or partially disordered systems, the Czjzek or extended Czjzek model is typically employed to construct the corresponding NMR line. The Czjzek model allows one to determine the quadrupolar parameters influencing the shape of the NMR spectra under the assumption of a Gaussian distribution of the EFG values with zero mean values. On the other hand, the extended Czjzek model assumes that the EFG tensor can be expressed as: $V = V_0 + \varepsilon V_G$, where $V_0$ denotes a well-defined contribution arising from the average local symmetry, and $V_G$ is a Gaussian-distributed random component described by the Czjzek model. The parameter ε controls the degree of admixture between the ordered and disordered contributions to the EFG [56, 57].

In this context, the inset in **Fig. 6** shows the fit of the experimental spectrum, where the regions with preserved average local symmetry are modeled using the extended Czjzek approach. A slightly larger quadrupolar constant is obtained, $C_Q$ = 19 MHz, with the best fit corresponding to η = 0.8, consistent with orthorhombic symmetry. For comparison, the line shape associated with a tetragonal phase is characterized by $C_Q$ = 19 MHz and η = 0 under the extended Czjzek distribution; it is shown at the bottom of the inset in **Fig. 6**.

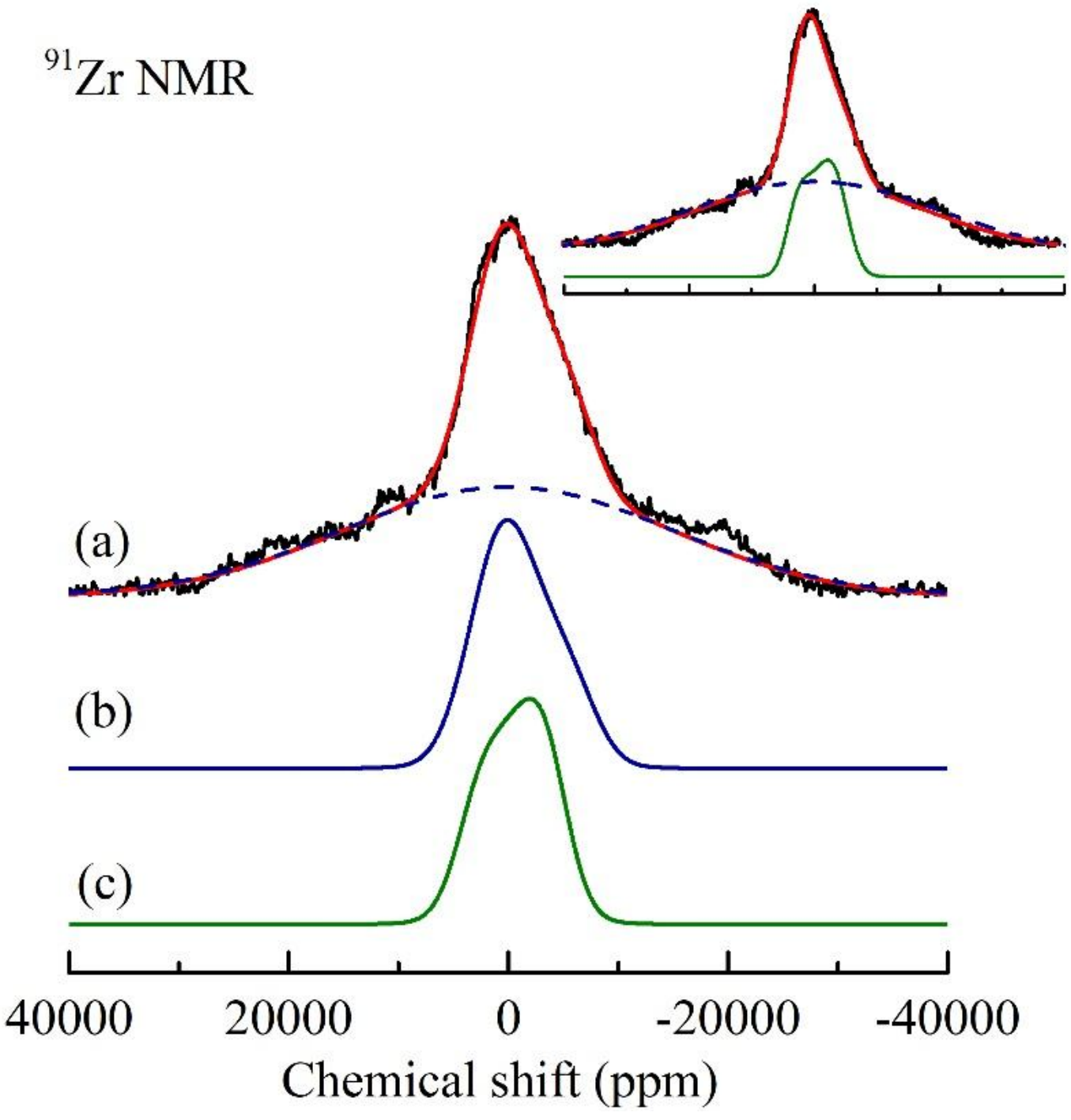

**Figure 6. (a)** Experimental $^{91}$Zr NMR spectrum of $Hf_{0.5}Zr_{0.5}O_2$ nanoparticles annealed in a CO+$CO_2$ atmosphere along with the best fit, where the dashed curve presenting a Gaussian component; **(b)** the line shape corresponding to orthorhombic symmetry; and **(c)** the line shape corresponding to tetragonal symmetry. Inset: the experimental spectrum and its fits obtained using the extended Czjzek model ($\eta$ = 0.8, $\varepsilon$ = 0.3, red line), and the green curve obtained with the model assuming that $\eta$ = 0.0 and $\varepsilon$ = 0.3.

Analysis of the experimental spectra therefore indicates that the nanoparticles are stabilized predominantly in the o-phases, which can include nonpolar o-I, antipolar o-II, and polar o-III phases, and can be clearly distinguished from the tetragonal line shape by the asymmetry parameter of the EFG. However, considering that the nonpolar and antipolar o-phases can be regarded as mirror-reflected arrangements of monoclinic and polar phases, respectively [58], and taking into account that in the monoclinic phase the symmetry of the local environment is characterized by a value of $\eta$ close to 0, the NMR results suggest that the experimental compound is more consistent with either an antipolar or a polar o-phase. Despite the fact that the NMR study (as well as EPR and XRD) does not allow us to separate all three o-phases, theoretical modeling, presented in the next section, indicates the appearance and stability of the FE o-III phase in the presence of oxygen vacancies.

## III. THEORETICAL MODELING

### A. Model Description

Elastic strains can induce a ferroelectric state in $HfO_2$ thin films [15, 59, 60]. Recently, it has been shown theoretically that compressive chemical stress can induce a FE o-III phase and related polar properties in spherical $HfO_2$ nanoparticles [61]. Similar effects may be expected in the oxygen-deficient $Hf_{0.5}Zr_{0.5}O_2$ nanoparticles considered in this work. This expectation is based on the fact that oxygen vacancies can induce strong chemical strains in oxide ferroelectrics [62, 63].

To study this possibility theoretically, let us consider an ensemble of $Hf_{0.5}Zr_{0.5}O_2$ core-shell nanoparticles with a spherical shape and narrow size distribution. The average particle radius is $R \approx 3.5$ nm with a standard deviation of the particle size distribution of $\delta \approx 0.5$ nm is obtained from TEM measurements. The cores are assumed to be defect-free, crystalline, and insulating. The core is covered with a thin shell, whose thickness $\Delta R$ is much smaller than the core radius $R_c$ (see **Fig. 7**). The shell is assumed to be semiconducting due to the high concentration of oxygen vacancies, which can be ionized. Both neutral and ionized vacancies act as elastic defects, which induce strong chemical strains with a magnitude $w_s$. These strains are proportional to the product of the Vegard tensor magnitude $W_s$ and the defect concentration $n$, i.e., $w_s \cong W_s n$. The absolute value of $|W_s|$ (also referred to as the "elastic dipole") is about $3 - 30$ Å$^3$ in oxide ferroelectrics [62, 63]. Therefore, the estimation $n \approx (0.3 - 3)\cdot 10^{27}$ m$^{-3}$ for the atomic concentration $n_0 \cong 6 \cdot 10^{28}$ m$^{-3}$ agrees with relevant experiments [64, 65, 66]. These experiments show that the defect concentration can exceed 0.5 – 5 % near the surface of various oxides, and that the corresponding chemical strain $w_s$ can reach several percent. Due to the elastic mismatch at the core-shell interface, the chemical strain induces elastic stress in the core [61]. The nanoparticles assumed to be placed in a dielectric medium, whose effective dielectric permittivity $\varepsilon_{eff}$ depends on their concentration. Following Ref. [61], we assume that the effective screening length $\lambda_{eff}$ in the shell depends on the defect concentration $n$ according to $\lambda_{eff} = \sqrt{\frac{\varepsilon_0 k_B T}{2e^2 n \varepsilon_{eff}}}$, where $\varepsilon_0$ is the vacuum dielectric constant, $k_B$ is the Boltzmann constant, $T$ is the absolute temperature, and $e$ is the elementary charge.

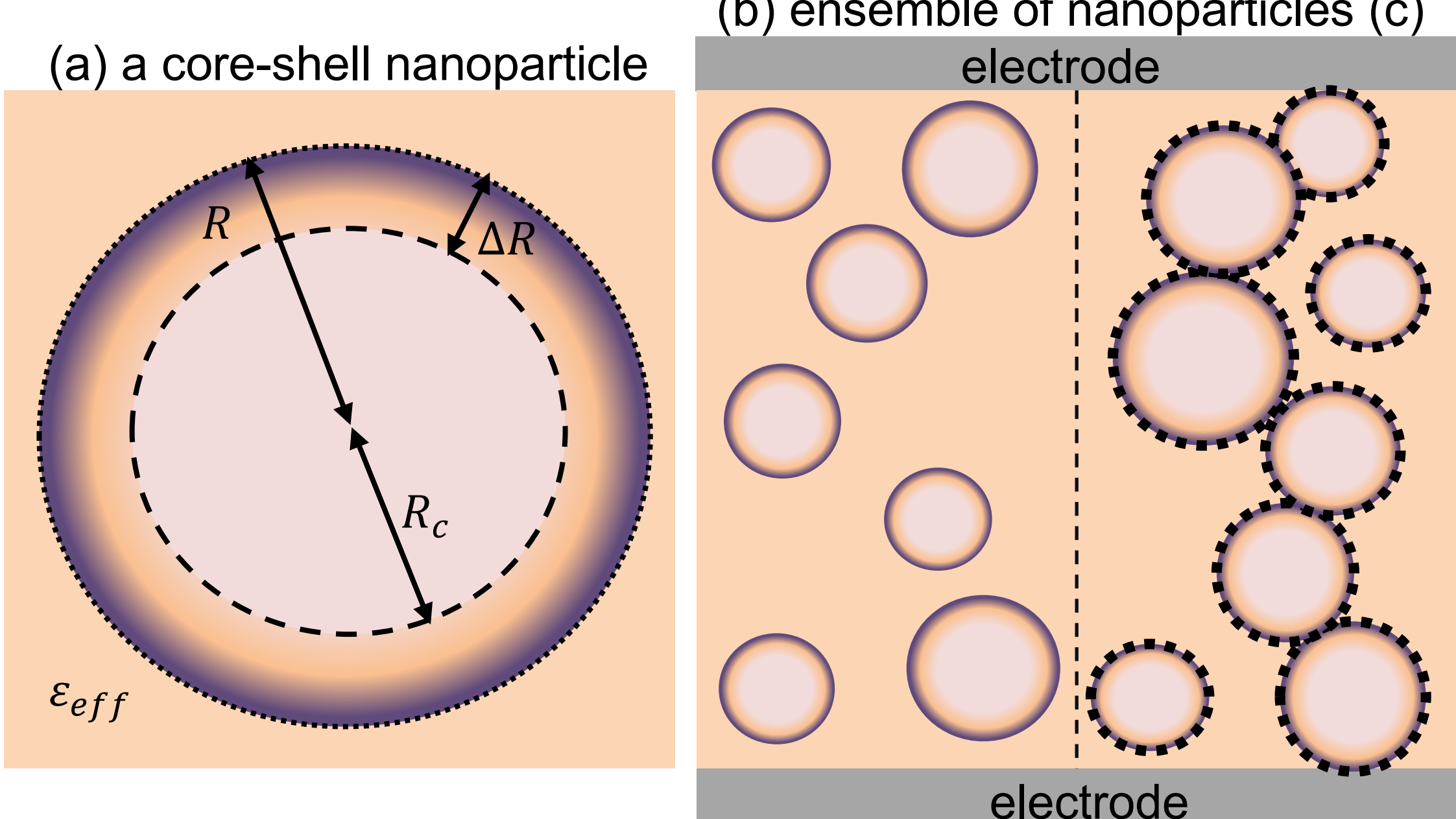

**Figure 7. (a)** The cross-section of a spherical $Hf_{0.5}Zr_{0.5}O_2$ core-shell nanoparticle: a defect-free core of radius $R_c$ is covered with a paraelectric shell of thickness $\Delta R$, which is filled with elastic defects (oxygen vacancies) and free charges. **(b)** An ensemble of $Hf_{0.5}Zr_{0.5}O_2$ core-shell nanoparticles with an effective dielectric permittivity $\varepsilon_{eff}$. **(c)** An ensemble of clustered and well-screened $Hf_{0.5}Zr_{0.5}O_2$ core-shell nanoparticles. Part (a) is adapted from Ref. [61].

To describe the phase diagrams of the $Hf_{0.5}Zr_{0.5}O_2$ nanoparticles, we use the Landau-Ginzburg-Devonshire (LGD) approach proposed in Ref. [61]. The form of the free energy functional is based on the work of Delodovici et al. [67, 68], Lee et al. [69] and Jung and Birol [70, 71], where the principal role of the trilinear coupling between the polar, antipolar, and nonpolar modes has been established. According to these studies, the polar mode cannot appear alone (see e.g., Fig. 1 in Ref. [69]) since it should be accompanied by the antipolar and nonpolar modes, and the trilinear coupling of the modes leads to the trigger-type FE transition to the o-III phase. The trilinear coupling may explain why it is difficult to stabilize a purely polar o-III phase in the $Hf_{0.5}Zr_{0.5}O_2$ nanoparticles, while mixtures of polar, antipolar, and nonpolar o-phases can be observed.

The LGD free energy of the $Hf_{0.5}Zr_{0.5}O_2$ nanoparticle core corresponding to the FE o-III phase consists of the bulk energy density $f_{bulk}$, the electric energy $f_{el}$, and the surface energy $F_S$ [61]:

$$F_{o-phase} = \int (f_{bulk} + f_{el}) dV + F_S, \quad f_{bulk} = f_{bq} + f_{tr} + f_{est} + f_{grad}. \qquad (3)$$

The bulk energy density $f_{bulk}$ is the sum of the biquadratic energy $f_{bq}$ and trilinear coupling energy $f_{tr}$ of the polar, antipolar, and nonpolar order parameters, the elastic and striction energy contributions $f_{est}$, and the gradient energy of the order parameters $f_{grad}$. The energy $f_{bq}$ is an expansion over the even powers and $f_{tr}$ is an expansion over the odd powers of the dimensionless amplitudes $Q_{\Gamma 3}$, $Q_{Y2}$, and $Q_{Y4}$ of the polar phonon mode $\Gamma_{3-}$, nonpolar phonon mode $Y_{2+}$, and antipolar phonon mode $Y_{4-}$ (see Refs.[67, 68, 70, 71] for details). The energies $f_{bq}$, $f_{tr}$, $f_{est}$, $f_{grad}$, and $f_{el}$ are listed in Ref. [61]. Corresponding material parameters are listed in Ref. [21]. Assuming natural boundary conditions at the core-shell interface, namely $\frac{\partial Q_i}{\partial \vec{n}} = 0$, the surface energy $F_s$ can be neglected in Eq.(3). Following the arguments given in Refs. [60, 61, 68], we consider the cubic $C_{cce}$ phase as the parent phase of the $Hf_{0.5}Zr_{0.5}O_2$ core.

The polarization $P_3$ is proportional to the amplitude $Q_{\Gamma 3}$ of the $\Gamma_{3-}$ mode [60, 68, 70]:

$$P_3 = \frac{Z_B^* d}{V_{f.u.}} Q_{\Gamma 3} \approx P_0 Q_{\Gamma 3}, \qquad (4)$$

where the polarization amplitude $P_0 \approx 54.8$ μC/cm$^2$ at room temperature [68].

The ferroelectric state of the core can be stable when its free energy $F_{o-phase}$ is smaller than the energy of the m-phase, $F_m = \int f_m dV$, where $f_m$ is the energy density of the m-phase in bulk $Hf_{0.5}Zr_{0.5}O_2$.

### B. Theoretical Results

The averaged spontaneous polarization $\bar{P}_s$ and the antipolar order parameter $\bar{Q}_{Y4s}$ of $Hf_{0.5}Zr_{0.5}O_2$ nanoparticles as a function of defect concentration $n$ and elastic dipole $W_s$ are shown in **Figs. 8(a)** and **8(b)**, respectively. The color scales in **Figs. 8(a)** and **8(b)** represent the absolute values of $\bar{P}_s$ and $\bar{Q}_{Y4s}$ in the deepest potential well of the free energy (3). The sharp boundary between the ferroelectric o-phase (with $\bar{P}_s > 0$ and $\bar{Q}_{Y4s} > 0$) and nonpolar m-phase (with $\bar{P}_s = 0$ and $\bar{Q}_{Y4s} = 0$) is the first order phase transition curve describing the dependence of the critical elastic dipole $W_{cr}$ on the defect concentration $n$ at room temperature $T = 293$ K. As it was shown in Refs. [21, 61], the difference of the antipolar and nonpolar order parameters, $|\bar{Q}_{Y4s}| - |\bar{Q}_{Y2s}|$, is very small due to the negligible anisotropy of the deepest potential well as a function of $|\bar{Q}_{Y4s}|$ and $|\bar{Q}_{Y2s}|$. Therefore, the dependence of $|\bar{Q}_{Y2s}|$ versus $n$ and $W_s$ looks very similar to those of $|\bar{Q}_{Y4s}|$ and is not shown in **Fig. 8**.

**Figure 8** also shows that the FE o-III phase can be stable only under compressive chemical strains, namely at $W_s < W_{min}$, where $W_{min} \approx -(3-5)$ Å$^3$ for $\varepsilon_{eff} = 10 - 30$ and $\bar{R} = 3.5$ nm. Note that the values of $\varepsilon_{eff}$ used here agree with the experimentally measured values of the relative dielectric permittivity of $Hf_{0.5}Zr_{0.5}O_2$ nanopowders at high frequency [27]. The requirement of compressive chemical strains to induce the FE o-III phase in the oxygen-deficient $Hf_{0.5}Zr_{0.5}O_2$ nanoparticles agrees with DFT results for compressed $HfO_2$ thin films [59, 68]. The minimal concentration of defects, $n_{min}$, decreases with increasing $\varepsilon_{eff}$, from 1.2% at $\varepsilon_{eff} = 10$ to 0.8 % at $\varepsilon_{eff} = 30$. The stability region of the nonpolar m-phase decreases strongly with increasing $n$ (as well as with an increasing magnitude of the elastic dipole $W_s$) and is replaced by the FE o-III phase. The decrease of the m-phase stability region and the minimal concentration $n_{min}$ is explained by the increase of dielectric and space-charge screening as $\varepsilon_{eff}$ increases [61].

The activation field $E_{af}$ of polarization reversal can be estimated as $E_{af} \cong b_{af}/\bar{P}_s$, where $b_{af}$ is the lowest barrier of polarization switching (see details in Refs. [60, 61]). The value of $b_{af}$ is nearly constant, $b_{af} \cong 48$ meV/f.u., and the average spontaneous polarization $\bar{P}_s$ depends on the elastic dipole $W_s$, defect concentration $n$, nanoparticle radius $R$, and shell thickness $\Delta R$ [61].

Color maps of $E_{af}$ as a function of $W_s$ and $n$ are shown in **Figs. 8(e)** and **8(f)**. They are calculated for two values of the effective permittivity, $\varepsilon_{eff} = 10$ and 30. The values of $E_{af}$ change from 0.68 MV/cm (far from the o-m phase boundary) to 0.98 MV/cm (near the o-m phase boundary) and are very close to the activation fields calculated by us earlier in $HfO_2$ thin films [60] and nanoparticles [61]. These activation fields are lower than the coercive fields $E_c \cong 1.05 - 1.35$ MV/cm observed experimentally in 10-nm thick $Hf_{0.5}Zr_{0.5}O_2$ films [72], because the local nucleation of nanodomains at $E_{af}$ precedes the global polarization switching at $E_c$. The region of the lowest $E_{af}$ corresponds to the largest $n$ and $W_s$, because the compressive chemical stress is largest in this region. The region of the lowest $E_{af}$ increases with increasing $\varepsilon_{eff}$. The increase occurs due to a decrease in the depolarization field, as shown in Ref. [61].

Color maps of the linear relative dielectric permittivity of the nanoparticle core, $\varepsilon_c \cong \frac{1}{\varepsilon_0}\frac{dP_3}{dE_3} + \varepsilon_b$, as a function of $W_c$ and $n$ are shown in **Figs. 8(g) – 8(h)**, for $\varepsilon_{eff} = 10$ and 30. As shown in **Fig. 8**, the magnitude of $\varepsilon_c$ can be tuned by the choice of $W_c$ and/or $n$ from 10 to 35 at room temperature. The permittivity reaches maximal values (a little over 30) in a region offset from the boundary between the o- and m-phases. The region of maximal $\varepsilon_c$ has the shape of a wide curved stripe and can be attributed to a significant contribution of the nonpolar and antipolar orders to the permittivity due to the trilinear coupling in the free energy (3). The permittivity does not diverge at the o-m boundary because the transition is first order.

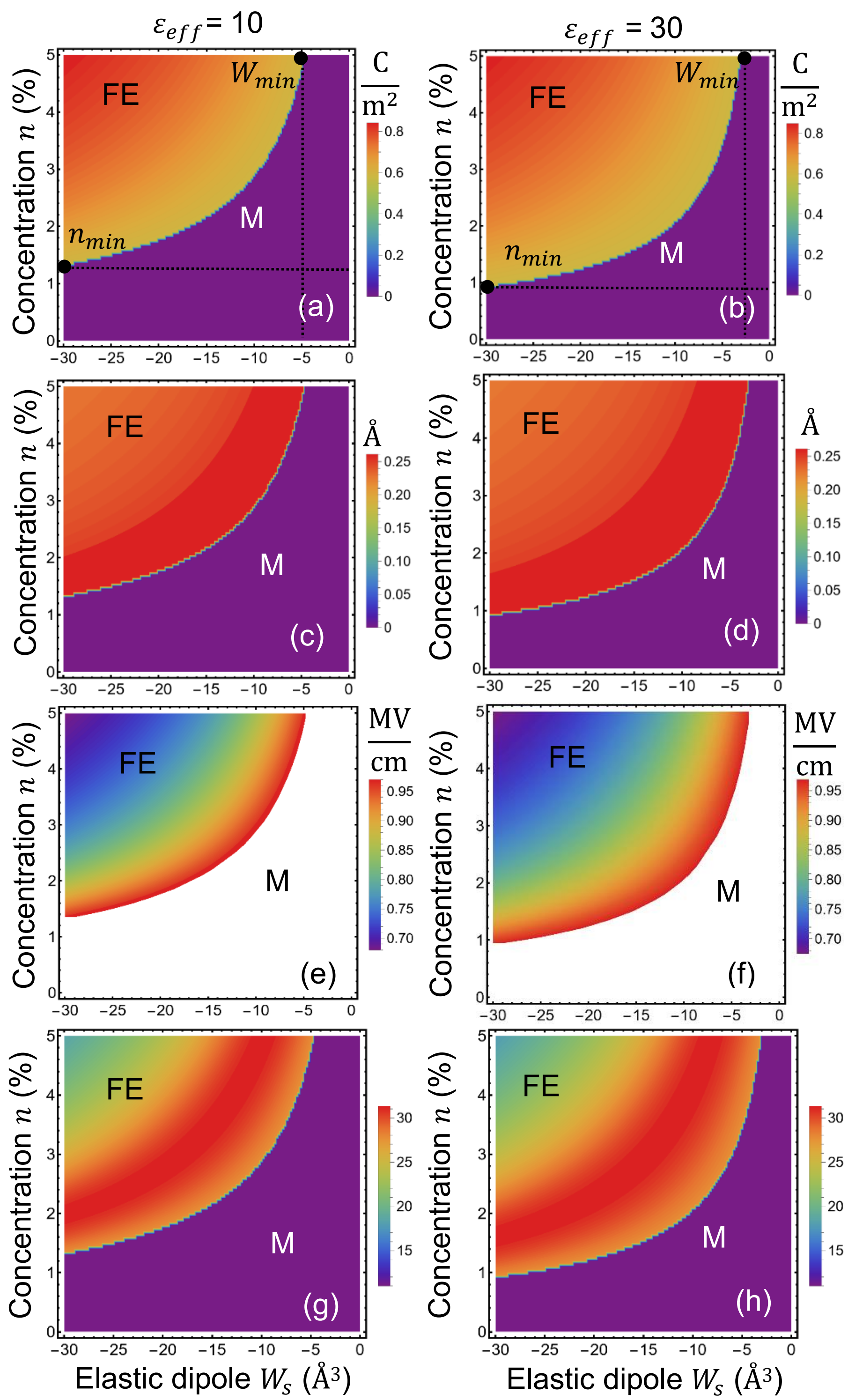

**Figure 8.** The absolute value of the spontaneous polarization $\bar{P}_S$ **(a, b)**, the amplitude of the antipolar order parameter $|\bar{Q}_{Y4s}|$ **(c, d)**, the activation field $E_{af}$ **(e, f)**, and the local dielectric permittivity of the core $\varepsilon_c$ **(g, h)** as a function of the Vegard strain $W_s$ and the defect concentration $n$, calculated for effective permittivities $\varepsilon_{eff} = 10$ **(a, c, e, g)** and 30 **(b, d, f, h)**. The abbreviation "FE" denotes the FE o-III phase, and "M" denotes the nonpolar m-phase of the $Hf_{0.5}Zr_{0.5}O_2$ nanoparticles. The average radius of the nanoparticles is $\bar{R} = 3.5$ nm, the shell thickness $\Delta R = 1.5$ nm, and the temperature is $T = 293$ K. Material parameters used in the calculations are listed in **Tables S1-S2** in **Appendix S1** of Ref. [60].

The calculated values of $\varepsilon_c \approx 10 - 30$ agree well with the effective permittivity $\varepsilon_{eff}$ measured at 100 – 500 kHz in the densely pressed oxygen-deficient $Hf_xZr_{1-x}O_{2-y}$ nanopowders (x = 1, 0.6, 0.5, and 0.4, and $0 < y << 1$) with the 7 – 10 nm size of the particles [27]. The nanoparticles studied in Ref. [27] were annealed in a $CO+CO_2$ atmosphere to induce a high concentration of oxygen vacancies near the surface. Thus, calculated $\varepsilon_c$ values are of the same order of magnitude as experimentally measured permittivity and are consistent with the experimental results in Ref. [27].

### C. Effective Dielectric Permittivity of the $Hf_{0.5}Zr_{0.5}O_2$ – PVDF Nanocomposites: Experiment and Theory

Experimental measurements of the nanopowders dielectric permittivity have a serious drawback related with the possible redistribution of the nanoparticles during the measurement, and thus the obtained results are highly sensitive to the applied pressure in the powder cell. Therefore, it is reasonable to measure and analyze the effective dielectric permittivity of the composites consisting of $Hf_{0.5}Zr_{0.5}O_2$ nanoparticles homogeneously distributed and fixed in a PVDF matrix. The studied nanocomposites $Hf_{0.5}Zr_{0.5}O_2$ – PVDF contain $\mu$ =13 vol. % of the $Hf_{0.5}Zr_{0.5}O_2$ nanoparticles annealed in air (sample N1) or annealed in $CO+CO_2$ (sample N2). The effective dielectric permittivity of the nanocomposite is a complex value, namely $\varepsilon^*_{eff} = \varepsilon_{eff} + i\frac{\sigma}{\varepsilon_0\omega}$, where $\varepsilon_{eff}$ is the real part of the permittivity, and the imaginary part of the permittivity is proportional to the electrical conductivity $\sigma$ determining the effective dielectric losses at frequency $\omega$. Our goal is to relate the values of $\varepsilon_c$, calculated in the previous section, to the values that can be determined (e.g., extracted or deconvoluted) from the measured values of the composite effective dielectric permittivity $\varepsilon^*_{eff}$.

Temperature dependences of the $\varepsilon^*_{eff}$, measured in the frequency range from ~100 Hz to 1 MHz, are shown in **Fig. 9**. The permittivity $\varepsilon_{eff}$ of the PVDF plates with $Hf_{0.5}Zr_{0.5}O_2$ nanoparticles annealed in air (sample N1) is smaller than that of the plates containing $Hf_{0.5}Zr_{0.5}O_2$ nanoparticles annealed in the $CO+CO_2$ ambient (sample N2) (compare **Fig. 9(a)**) and **Fig. 9(b)**). At low frequencies, the temperature dependence of $\varepsilon_{eff}$ has two distinct maxima located at approximately 350 K and 430 K. Broadening and overlap of the two features are observed as the concentration of oxygen vacancies increases, causing the lower-temperature maximum to evolve into a shoulder (compare **Fig. 9(a)** with **Fig. 9(b)**). The magnitude of $\varepsilon_{eff}$ decreases monotonically with an increase in frequency. Both features of $\varepsilon_{eff}$ exhibit a pronounced dispersion in their intensity, decreasing strongly and vanishing as the frequency increases from 170 Hz to 1 MHz (only weak shoulder-like features remain in the dark-blue curves corresponding to 1 MHz in **Figs. 9(a)** and **9(b)**).

The effective dielectric losses, shown in **Fig. 9(c)** and **Fig. 9(d)**, are very large at low frequencies (for which $\mathrm{Re}[\varepsilon_{eff}^{*}] \ll \mathrm{Im}[\varepsilon_{eff}^{*}]$) and much smaller at high frequencies (for which $\mathrm{Re}[\varepsilon_{eff}^{*}] \gg \mathrm{Im}[\varepsilon_{eff}^{*}]$). The corresponding loss tangent, namely $\tan\delta$ (where $\delta = \frac{\mathrm{Im}[\varepsilon_{eff}^{*}]}{\mathrm{Re}[\varepsilon_{eff}^{*}]}$), is shown in **Fig. 9(e)** and **Fig. 9(f)**, respectively. The values of $\mathrm{Im}[\varepsilon_{eff}^{*}]$ and $\tan\delta$ increase with increasing temperature at low frequencies (less than 50 kHz) and decrease with increasing temperature at high frequencies (more than 100 kHz). The behavior of the losses is relatively common for polymer composites with a high volume content of ferroelectric nanoparticles (namely more than 10 %) (see e.g., Ref. [73] and references therein). The increase of losses is likely associated with thermally activated ionic transport and/or hopping conduction mechanisms [27, 28, 73]; in this case, the interfaces and shells of the nanoparticles are likely the main source of mobile charges. The losses decrease monotonically with increasing frequency from 100 Hz to 1 MHz for both types of nanocomposites, dropping by 2-3 orders of magnitude over this range, because the transport of screening charges is sluggish in the weakly-conducting polymer PVDF.

Such behavior of $\varepsilon_{eff}^{*}$ maybe attributed to the Maxwell-Wagner (MW) effect, which arises due to the formation of volume charges at the interfaces between different materials (nanoparticles and polymer in the considered case). However, as it was shown in Refs. [74, 75, 76], the MW-type of effective dielectric permittivity reaches colossal values (more than $10^{4}$) at low frequencies. This magnitude of permittivity of nanograined ferroelectric ceramics [77], nanopowders [27], and/or nanocomposites is typically caused by mesoscopic inhomogeneities in electrical conductivity (mainly between grains/particles and their boundaries), known as internal barrier layer capacitance (IBLC). Additional contributions may arise from inhomogeneous layers between electrodes and the sample, known as surface barrier layer capacitance (SBLC). The IBLC and SBLS effects produce a colossal dielectric response at low frequencies when the conductivity of the material between grain (or particle) boundaries is significantly lower than that of the grains (or particles). In this case, percolation of components with low conductivity occurs. This effect can be described by the effective medium approach (EMA), as discussed in Refs. [78, 79], as well as in the works of Petzelt et al. [80] and Richetsky et al. [81].

The colossal dielectric permittivity is not observed experimentally in the studied nanocomposites. The maximal value of $\varepsilon_{eff}$ is about 23 in **Fig. 9(a)** and does not exceed 45 in **Fig. 9(b).** The dielectric permittivity of pure PVDF varies from 10 – 12 at low frequencies to 5 – 7 at high frequencies [73]. Using EMA, including the Maxwell-Garnet and Bruggeman approximations [82], we estimate the possible contributions of MW-type effects and ferroelectric-like dipolar contributions to the effective dielectric permittivity.

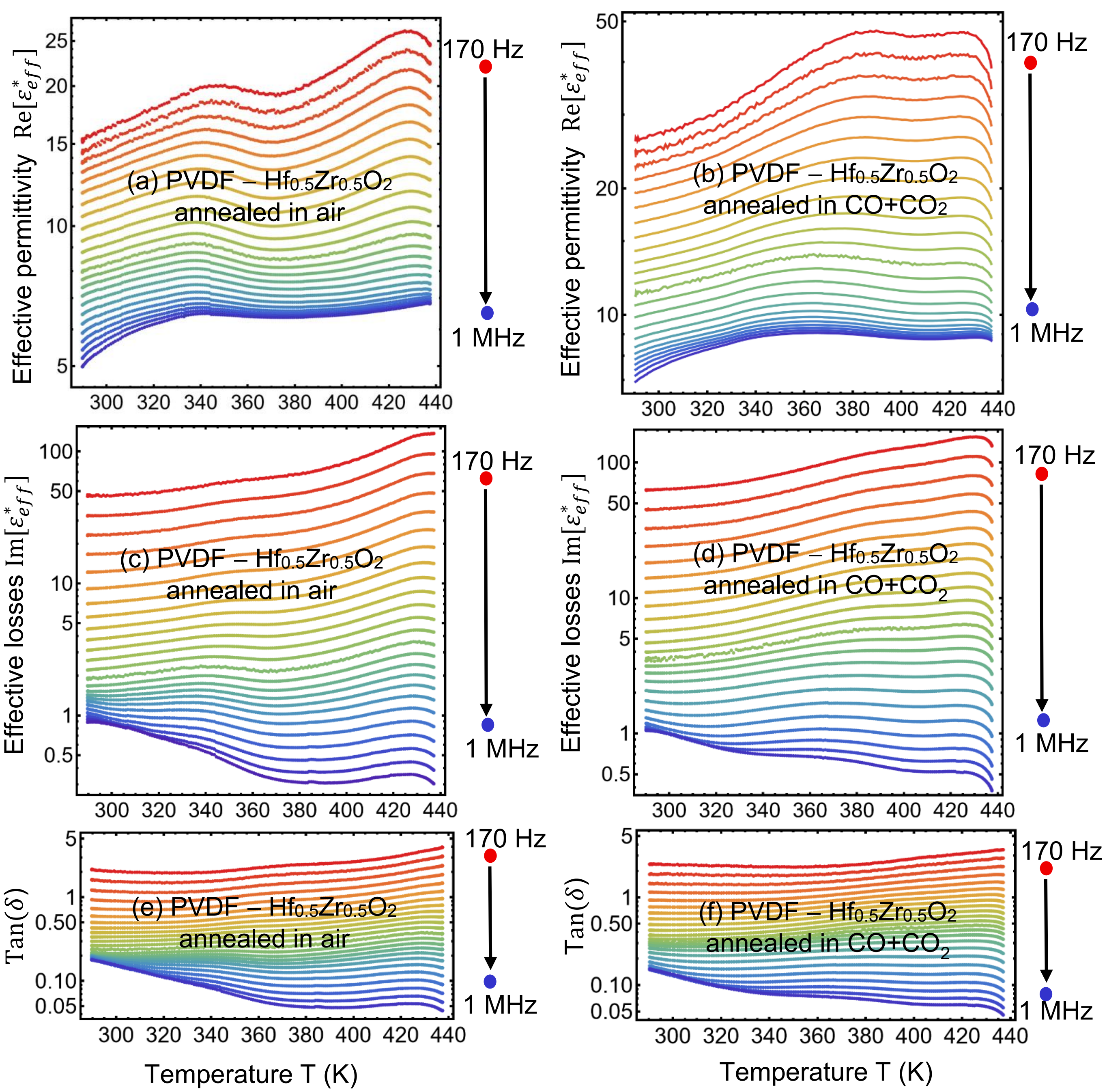

**Figure 9**. Temperature dependences of the real **(a, b)** and imaginary **(c, d)** parts of the effective permittivity $\varepsilon^*_{eff}$, and the loss tangent tan$\delta$ **(e, f)** of $Hf_{0.5}Zr_{0.5}O_2$ – PVDF nanocomposites measured in the frequency range from 170 Hz (top red curves) to 1 MHz (bottom dark-blue curves). The $Hf_{0.5}Zr_{0.5}O_2$ nanopowders used for the nanocomposite preparation were annealed in air, sample N1 **(a, c, e)**, or in a CO+$CO_2$ atmosphere, sample N2 **(b, d, f)**.

Both the minimal ($\varepsilon_{eff} \approx 25$) and maximal ($\varepsilon_{eff} \approx 43$) values of $\varepsilon_{eff}$ measured at 170 Hz for sample N2 (annealed in CO+$CO_2$) may be described by the Maxwell-Garnet or Bruggeman approximations [83] for $\mu = 13$ vol. %, but only if one assume large values of the nanoparticle permittivity, $\varepsilon_{NP} > 10^2 - 10^4$ (see **Supplement S2** for details). These values of $\varepsilon_{NP} \sim 10^3 - 10^4$ are significantly larger than the values of $\varepsilon_c$ calculated using LGD approach and shown in **Fig. 8**. This discrepancy may indicate the emergence of a superparaelectric (SPE)-like state in the $Hf_{0.5}Zr_{0.5}O_2$ nanoparticles annealed in $CO+CO_2$. The SPE-like state often precedes the negative capacitance (NC) states of the ferroelectric nanoparticles, which may exist in the "dense" fine-grained ceramics [84] and composites with a volume fraction of nanoparticles $\mu > 0.1$ [73, 77, 85].

Thus, the discrepancy between the values of $\varepsilon_c$ calculated using the LGD approach and the values of $\varepsilon_{NP}$ estimated from experiment could indicate additional mechanisms contributing to the observed increase in $\varepsilon_{eff}$, including relaxor-like behavior, SPE-like behavior, or NC states of the $Hf_{0.5}Zr_{0.5}O_2$ nanoparticles. A possible interplay of MW-type, IBLS, and SBLS effects; dipolar polarization; and SPE-like and NC states can be analyzed using the EMA, which yields an algebraic equation for determining the complex dielectric permittivity $\varepsilon^*_{eff}$ of the effective medium [77, 86]:

$$(1-\mu)\frac{\varepsilon^*_{eff}-\varepsilon^*_p}{(1-\eta_{NP})\varepsilon^*_{eff}+\eta_{NP}\,\varepsilon^*_p}+\mu\frac{\varepsilon^*_{eff}-\varepsilon^*_{NP}}{(1-\eta_{NP})\varepsilon^*_{eff}+\eta_{NP}\,\varepsilon^*_{NP}}=0. \qquad (5)$$

Here, $\varepsilon^*_{NP}$ is the complex relative permittivity of the polarized nanoparticles "*NP*", which are assumed to be monodisperse and uniformly polarized [73, 77]. The real part of $\varepsilon^*_{NP}$ corresponds to the dielectric (polarization) response, and the imaginary part of $\varepsilon^*_{NP}$ accounts for dielectric losses due to finite ionic conductivity, these losses give rise to MW-type and SBLS effects. $\varepsilon^*_p$ is the complex relative permittivity of the dielectric polymer "*p*", for which $\mathrm{Re}\left[\varepsilon^*_p\right] \gg \mathrm{Im}[\varepsilon^*_p]$. The values $\mu$ and $1-\mu$ are relative volume fractions of the components "*NP*" and "*p*", respectively. The maximal value $\mu_{max}$ is limited by the dense packing of the particles. In particular, $\mu_{max}=\frac{\pi}{6}\approx 0.52$ for densely packed nanospheres of a uniform radius. The function $n_{NP}$ is the depolarization field factor of the polarized nanoparticles, which is determined by their shape and polarization orientation; $0 \leq n_{NP} \leq 1$. Analytical expressions for $n_{NP}$ exist for uniformly polarized nanoparticles of ellipsoidal shape including nanowires, nanospheres, and nanodisks [87]. Screening charges existing in the nanoparticle shell weaken the depolarization field inside the particle, and the corresponding decrease of the "effective" depolarization factor $\eta_{eff}$ can be roughly estimated as $\frac{\eta_{NP}}{1+(R_c/\lambda_{eff})}$ [61]. For instance, $\eta_{eff} < 0.1\eta_{NP}$ for $\lambda_{eff} < 1$ nm and $R_c$ >10 nm.

The solution of Eq.(5) for identical nanoparticles, which are uniformly polarized (i.e., have an isotropic depolarization factor $\eta_{NP}$), is the following:

$$\varepsilon^{\pm}_{eff}=\frac{(1-\mu)\varepsilon^*_p+\mu\varepsilon^*_{NP}-\eta_{NP}\left(\varepsilon^*_p+\varepsilon^*_{NP}\right)\pm\sqrt{\left[(1-\mu)\varepsilon^*_p+\mu\varepsilon^*_{NP}-\eta_{NP}\left(\varepsilon^*_p+\varepsilon^*_{NP}\right)\right]^2+4\varepsilon^*_p\varepsilon^*_{NP}(1-\eta_{NP})\eta_{NP}}}{2(1-\eta_{NP})}. \qquad (6)$$

Since the PVDF matrix with embedded $Hf_{0.5}Zr_{0.5}O_2$ nanoparticles is the part of a capacitor structure in the dielectric measurements, both signs before the radical in Eq.(6) do not contradict thermal equilibrium. Since the sign "+" corresponds to a higher value of the effective permittivity, we consider this sign to correspond to the deepest energy minimum.

Assuming that $\eta_{NP} \to 0$, which corresponds to well-screened ($\lambda_{eff} \leq 0.1$ nm) nanoparticles and/or to the case where they form percolation clusters between the electrodes (see e.g., **Fig. 7(c)**), we obtain from Eq.(6) that $\varepsilon_{eff}^{*} = \varepsilon_{p}^{*}(1-\mu) + \mu\varepsilon_{NP}^{*}$. Assuming that $\eta_{NP} = 1/3$, which corresponds to spatially-isolated and weakly screened ($\lambda_{eff} \gg 1$ nm) spherical nanoparticles (see e.g., **Fig. 7(b)**), we obtain from Eq.(6) that

$$\varepsilon_{eff}^{\pm} = \frac{1}{4}\varepsilon_{p}^{*}(2-3\mu) + \frac{1}{4}\varepsilon_{NP}^{*}(3\mu-1) \pm \frac{1}{4}\sqrt{\left[\varepsilon_{p}^{*}(2-3\mu) + \varepsilon_{NP}^{*}(3\mu-1)\right]^{2} + 8\varepsilon_{p}^{*}\varepsilon_{NP}^{*}}. \qquad (7)$$

To analyze the experimental results, shown in **Fig. 9**, we need to extract the temperature dependence of $\varepsilon_{NP}^{*}$ from Eq.(5). The expression for $\varepsilon_{NP}^{*}$ has a relatively simple form:

$$\varepsilon_{NP}^{*} = \varepsilon_{eff}^{*}\frac{\mu\left[(1-\eta_{NP})\varepsilon_{eff}^{*}+\eta_{NP}\,\varepsilon_{p}^{*}\right]+(1-\eta_{NP})(1-\mu)\left(\varepsilon_{eff}^{*}-\varepsilon_{p}^{*}\right)}{\mu\left[(1-\eta_{NP})\varepsilon_{eff}^{*}+\eta_{NP}\,\varepsilon_{p}^{*}\right]-\eta_{NP}\,(1-\mu)\left(\varepsilon_{eff}^{*}-\varepsilon_{p}^{*}\right)} = \begin{cases} \dfrac{\varepsilon_{eff}^{*}}{\mu} - \varepsilon_{p}^{*}\dfrac{1-\mu}{\mu}, & \eta_{NP} \to 0, \\ \varepsilon_{eff}^{*}\dfrac{\mu\left[2\varepsilon_{eff}^{*}+\varepsilon_{p}^{*}\right]+2(1-\mu)\left(\varepsilon_{eff}^{*}-\varepsilon_{p}^{*}\right)}{\mu\left[2\varepsilon_{eff}^{*}+\varepsilon_{p}^{*}\right]-(1-\mu)\left(\varepsilon_{eff}^{*}-\varepsilon_{p}^{*}\right)}, & \eta_{NP} = \dfrac{1}{3}. \end{cases} \qquad (8)$$

It seems reasonable to assume that the polarized nanoparticles are electrically screened due to the finite screening length $\lambda_{eff}$ provided by ambient charges; some of them can be connected in long clusters, while others are spatially isolated (see e.g., **Fig.7(b) – 7(c)**). Thus, the factor $\eta_{NP}$ can be considered an "effective" fitting parameter that may vary from 0 (ideally screened particles) to 1/3 (spatially isolated unscreened particles). By considering that the real part of $\varepsilon_{p}^{*}$ decreases monotonically from 10 to 5 with increasing frequency from 100 Hz to 1 MHz [73], $\mu \approx 0.13$ and $\eta_{NP}$ can vary from 0 to 1/3, we can extract the complex dielectric permittivity $\varepsilon_{NP}^{*}$ of the $Hf_{0.5}Zr_{0.5}O_2$ nanoparticles from the experimental results shown in **Fig. 9** using Eq.(8).

The results for the two limiting cases, $\eta_{NP} = 0$ and $\eta_{NP} = 1/3$, are shown in **Figs. 10** and **11**, respectively. These results look completely different from those expected in the case of a dominating MW-type contribution, because the temperature dependences of $\mathrm{Re}[\varepsilon_{NP}^{*}]$ exhibit two pronounced and relatively well-separated maxima for the $Hf_{0.5}Zr_{0.5}O_2$ nanoparticles annealed in air, and a broader double-peaked maximum for the $Hf_{0.5}Zr_{0.5}O_2$ nanoparticles annealed in a $CO+CO_2$ atmosphere at low frequencies. The temperature dependences of $\mathrm{Im}[\varepsilon_{NP}^{*}]$ do not have any visible minima at low frequencies in the $Hf_{0.5}Zr_{0.5}O_2$ nanoparticles annealed in a $CO+CO_2$ atmosphere. In the case of a dominating MW-type contribution, the dielectric permittivity increases monotonically with temperature and does not have any pronounced maxima [74, 75]. Also, the maximum of the loss tangent corresponds to a shoulder or broad feature in the dielectric permittivity temperature dependence for an MW-type dielectric response [74, 75]. However, the temperature dependences of $\mathrm{Re}[\varepsilon_{NP}^{*}]$ do not obey Curie-Weiss type or similar laws. They resemble the dielectric response of disordered ferroelectrics with relaxor-like features.

It is seen from the comparison of **Figs. 10(a)** and **10(b)** that the contribution of dipolar polarization appears more significant in the $Hf_{0.5}Zr_{0.5}O_2$ nanoparticles annealed in a $CO+CO_2$ atmosphere. Consequently, the dielectric permittivity peak, observed near 350 K in the $Hf_{0.5}Zr_{0.5}O_2$ nanoparticles annealed in air (see **Fig. 10(a)**), can be related to oxygen vacancies. An increase in the vacancy concentration leads to a significant increase in the intensity of the peak, accompanied by a shift to higher temperatures and substantial broadening, as observed for the $Hf_{0.5}Zr_{0.5}O_2$ nanoparticles annealed in a $CO+CO_2$ atmosphere (see **Figs. 10(b)**). The vacancies lead to the formation of a defect-induced elastic dipole and to an increase in ionic conductivity, which decreases the depolarization field (due to the decrease in the effective screening length $\lambda_{eff}$) and may induce a relaxor-like (or SPE-like) phase transition in the vacancy-enriched $Hf_{0.5}Zr_{0.5}O_2$ nanoparticles. The transition is associated with a very diffuse double maximum (or shoulder-like feature) of $\varepsilon_{NP}$ located in the temperature range 370 – 430 K (see **Fig. 10(b)**). The maximal values of $\varepsilon_{NP}$ reach several hundred at low frequencies. The diffuse maximum of $\varepsilon_{NP}$ corresponds to high losses, which are weakly temperature-dependent (see **Fig. 10(d)** and **10(f)**). The high-frequency values of $\varepsilon_{NP}$ vary from 20 to 30 in agreement with the theoretical predictions shown in **Fig. 8**. The losses decrease strongly and monotonically with increasing frequency, and the loss tangent does not exceed 0.1 – 0.5 at 1 MHz and $T$ >350 K (see **Fig. 10(f)**). The relatively large high-frequency dielectric permittivity and small high-frequency conductivity of the core-shell $Hf_{0.5}Zr_{0.5}O_2$ nanoparticles annealed in a $CO+CO_2$ atmosphere agree with the theoretical predictions of their ferroelectric-like and/or relaxor-like states induced by the oxygen vacancies concentrated mainly in the shells.

It is seen from **Fig. 11**, calculated from **Fig. 9** at $\eta_{NP} = 1/3$, that the appearance of NC states is possible in the weakly screened $Hf_{0.5}Zr_{0.5}O_2$ nanoparticles. These states are more probable in the $Hf_{0.5}Zr_{0.5}O_2$ nanoparticles annealed in a $CO+CO_2$ atmosphere, where the frequency and temperature ranges of their possible existence are larger (compare **Fig. 11(a)** and **Fig. 11(b)**). The double maximum of the negative dielectric permittivity appears for both low and high frequencies in the vacancy-enriched $Hf_{0.5}Zr_{0.5}O_2$ nanoparticles (see **Fig. 11(b)**). The dielectric losses have two maxima at high frequencies in the nanoparticles (see **Fig. 11(d)**). The loss tangent diverges at the point where $\varepsilon_{NP}$ changes sign and differs in sign to the left and right of the divergence (see **Fig. 11(f)**). The positions of the loss maxima are close to the temperature maxima of the negative dielectric permittivity. This indicates the dipolar nature of the dielectric permittivity maxima observed in the $Hf_{0.5}Zr_{0.5}O_2$ nanoparticles annealed in a $CO+CO_2$ atmosphere, rather than the MW-type behavior of the dielectric response.

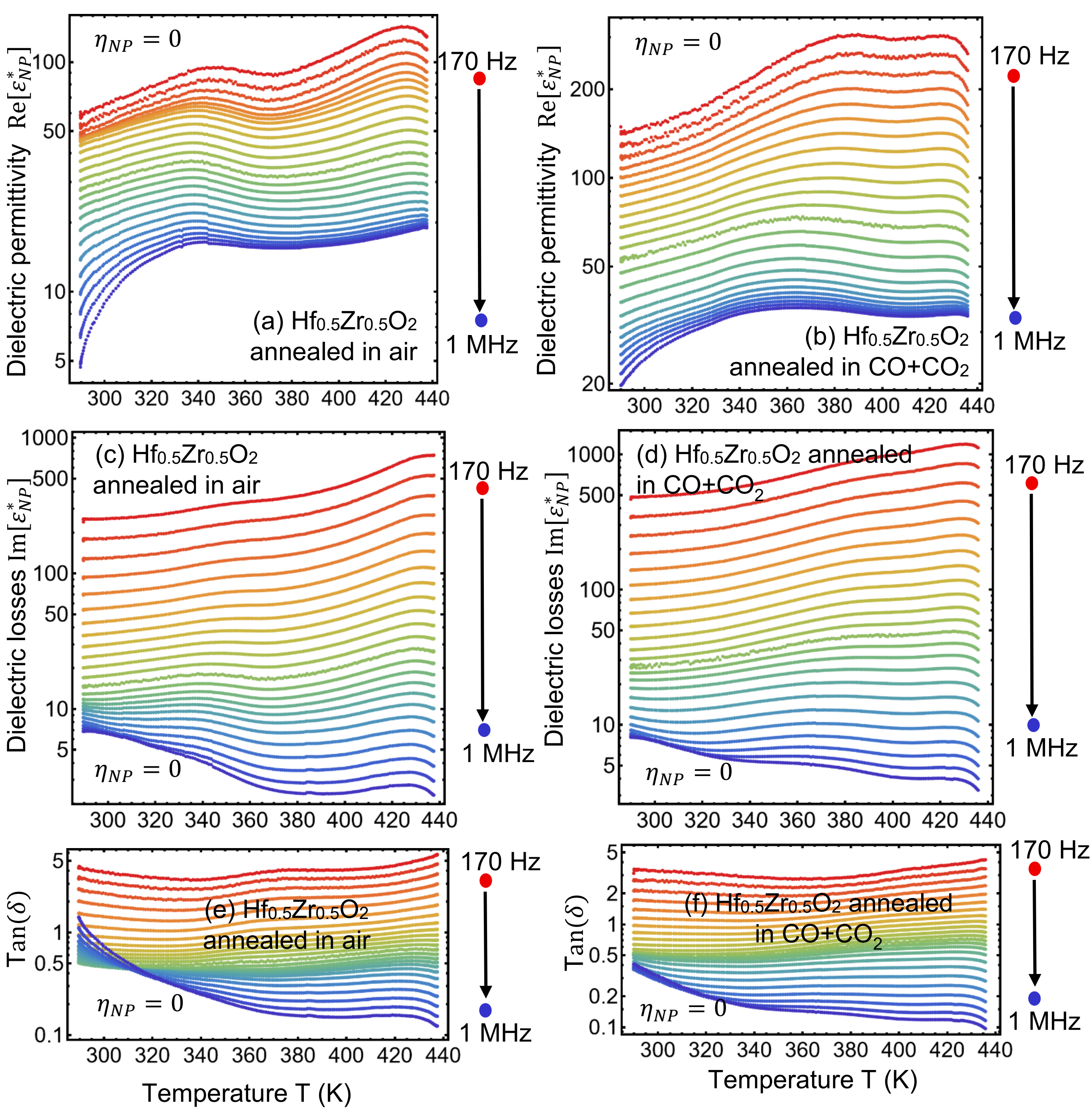

**Figure 10**. Temperature dependences of the real **(a, b)** and imaginary **(c, d)** parts of the dielectric permittivity $\varepsilon^*_{NP}$, and the loss tangent tan$\delta$ **(e, f)** of the $Hf_{0.5}Zr_{0.5}O_2$ nanoparticles. The frequency varies from 170 Hz (top red curves) to 1 MHz (bottom dark-blue curves). The curves are extracted from the experimentally measured $\varepsilon^*_{eff}$ at $\eta_{NP} = 0$. The $Hf_{0.5}Zr_{0.5}O_2$ nanopowders (N1 and N2) were annealed in air **(a, c, e)** and in $CO+CO_2$ atmospheres **(b, d, f)**.

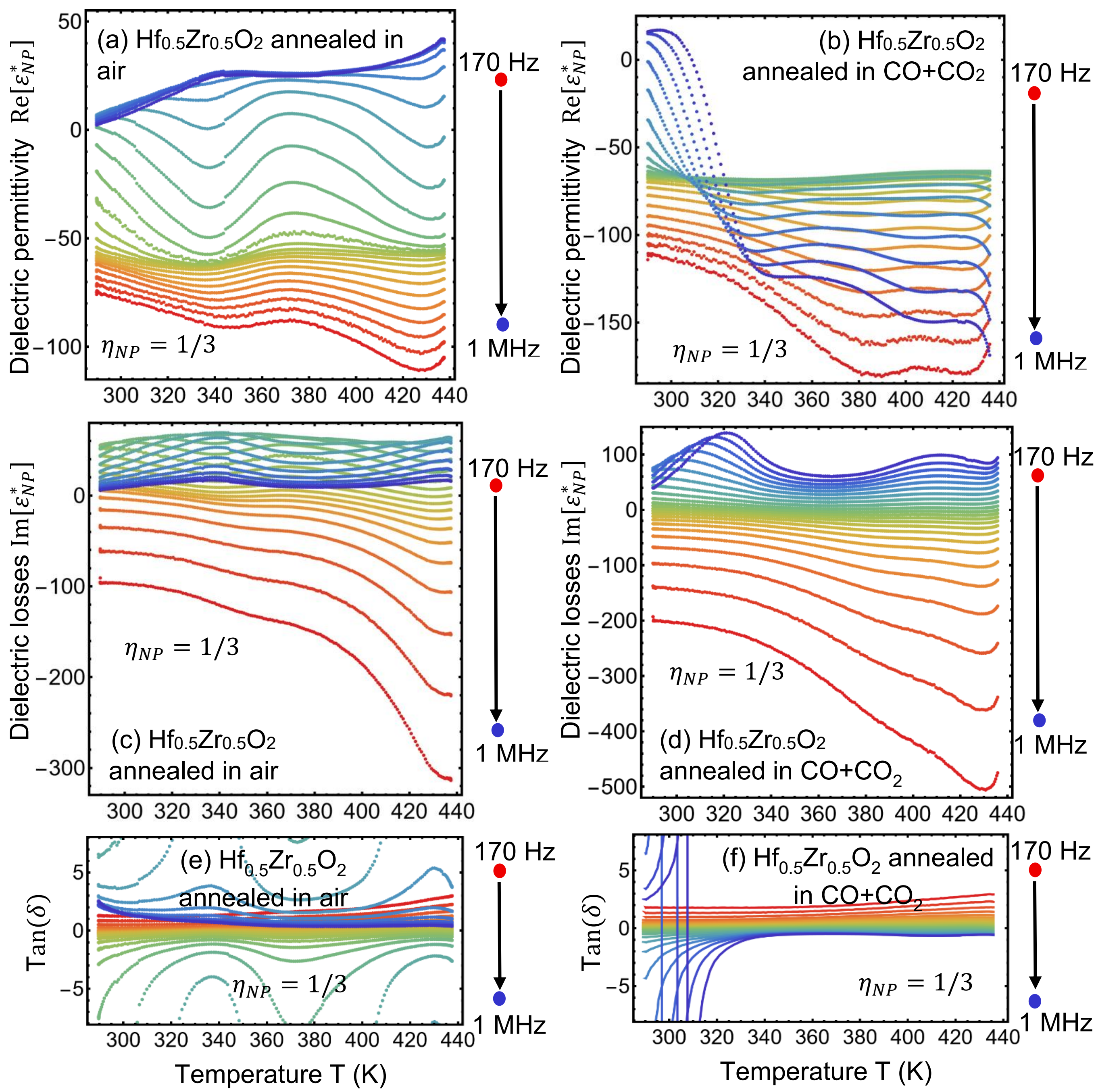

**Figure 11**. Temperature dependences of the real **(a, b)** and imaginary **(c, d)** parts of the dielectric permittivity $\varepsilon^*_{NP}$, and the loss tangent tan$\delta$ **(e, f)** of the $Hf_{0.5}Zr_{0.5}O_2$ nanoparticles. The frequency varies from 170 Hz (top red curves) to 1 MHz (bottom dark-blue curves). The curves are extracted from the experimentally measured $\varepsilon^*_{eff}$ at $\eta_{NP} = 1/3$. The $Hf_{0.5}Zr_{0.5}O_2$ nanopowders (N1 and N2) were annealed in air **(a, c, e)** and CO+$CO_2$ atmospheres **(b, d, f)**.

## IV. CONCLUSIONS

In this work, we study the stabilization of the orthorhombic phases in small $Hf_{0.5}Zr_{0.5}O_2$ nanoparticles (with an average size of 7 nm) annealed in air (oxidizing) or CO+$CO_2$ (reducing) atmospheres. The concentration of the oxygen vacancies, which is determined by the annealing conditions, was estimated to be as high as 10 – 15 % from EPR and XPS spectroscopy. The fraction of the orthorhombic phases (including nonpolar o-I, antipolar o-II, and polar o-III), as determined by X-ray diffraction and nuclear magnetic resonance, depends on the concentration of oxygen vacancies, which is controlled by the annealing conditions.

Phenomenological calculations based on Landau-Ginzburg-Devonshire theory confirm that the chemical strains induced by oxygen vacancies can stabilize the ferroelectric o-III phase with polar long-range ordering in small $Hf_{0.5}Zr_{0.5}O_2$ nanoparticles. We predict that the ferroelectric o-III phase can be stable in small core-shell $Hf_{0.5}Zr_{0.5}O_2$ nanoparticles at compressive chemical strains exceeding 1 – 5 % in the shell. The strains are well below the strains created by 10 – 15 % of oxygen vacancies, which concentration was estimated from XPS.

The contribution of dipole polarization was confirmed in the vacancy-enriched $Hf_{0.5}Zr_{0.5}O_2$ nanoparticles. Consequently, the increase in the intensity of the dielectric permittivity peak observed near 350 K in the PVDF film with the $Hf_{0.5}Zr_{0.5}O_2$ nanoparticles annealed in a $CO+CO_2$ atmosphere is clearly associated with an increase in the oxygen vacancy concentration. The vacancies lead to the formation of defect-induced elastic dipoles and to an increase in ionic conductivity, which reduces the depolarization field and may induce an SPE-like phase transition in the vacancy-enriched $Hf_{0.5}Zr_{0.5}O_2$ nanoparticles, as well as in other FE oxide nanoparticles. Due to the interfacial effects, the negative capacitance states may be realized in weakly screened and spatially isolated $Hf_{0.5}Zr_{0.5}O_2$ nanoparticles embedded in the PVDF matrix. More generally, the present approach based on oxygen-vacancy-induced elastic and screening effects may provide a route for engineering ferroelectric-like states in other nanoscale ferroic oxides.

## Authors' contribution

Y.O.Z. measured the EPR and NMR spectra, analyzed the data, evaluated the XPS results, and wrote the draft of the experimental part of the work. L.D. performed TEM, EDS, and EELS studies. V.V.L., P.J., and J.H. performed XPS measurements and analysis. O.V.L. and V.N.P. sintered the $Hf_xZr_{1-x}O_2$ nanopowders. A.O.D., M.D.V., and M.P.T. performed dielectric measurements. O.S.P. prepared electrode-coated samples for the measurements. M.V.K. performed XRD studies. E.A.E. performed numerical calculations and prepared corresponding figures. A.N.M. performed analytical calculations and wrote the theoretical part of the work. D.R.E. discussed and analyzed the results and made all improvements in the manuscript.

## Acknowledgments

The work of Y.O.Z., L.P.Y., A.O.D., M.D.V. and E.A.E. is funded by the National Research Foundation of Ukraine (grant N 2023.03/0127 "Silicon-compatible ferroelectric nanocomposites for electronics and sensors”). P.J. and J.H. acknowledge the support of the Strategy AV21 project: “Study of the atomically thin quantum materials by advanced microscopic/spectroscopic techniques applying machine learning”. L.D. acknowledges support from the Knut and Alice Wallenberg Foundation (grant no. 2018.0237) for TEM research. The work of O.V.L. and V.N.P. is supported by the National Academy of Sciences of Ukraine, III-6-26 “Innovative ferroelectric nanomaterials based on silicon-compatible oxides and nitrides of the rare earth and transition metals for strategic requirements of nanoelectronics”. Results of theoretical modelling were visualized in Mathematica 14.0 [88]. O.S.P. acknowledges the National Research Foundation of Ukraine (grant N 2023.03/0132 “Manyfold-degenerated metastable states of spontaneous polarization in nanoferroics: theory, experiment and perspectives for digital nanoelectronics”. A.N.M. acknowledges the support the EOARD project 9IOE063 (related STCU partner project is P751c) and (in the part of discussion and application of results) the National Research Foundation of Ukraine (grant N 2023.03/0132 “Manyfold-degenerated metastable states of spontaneous polarization in nanoferroics: theory, experiment and perspectives for digital nanoelectronics”) and the Horizon Europe Framework Programme (HORIZON-TMA-MSCA-SE), project № 101131229, Piezoelectricity in 2D-materials: materials, modeling, and applications (PIEZO 2D).

## SUPPLEMENTARY MATERIALS

### Supplement S1. Sample Characterization and Dielectric Measurements

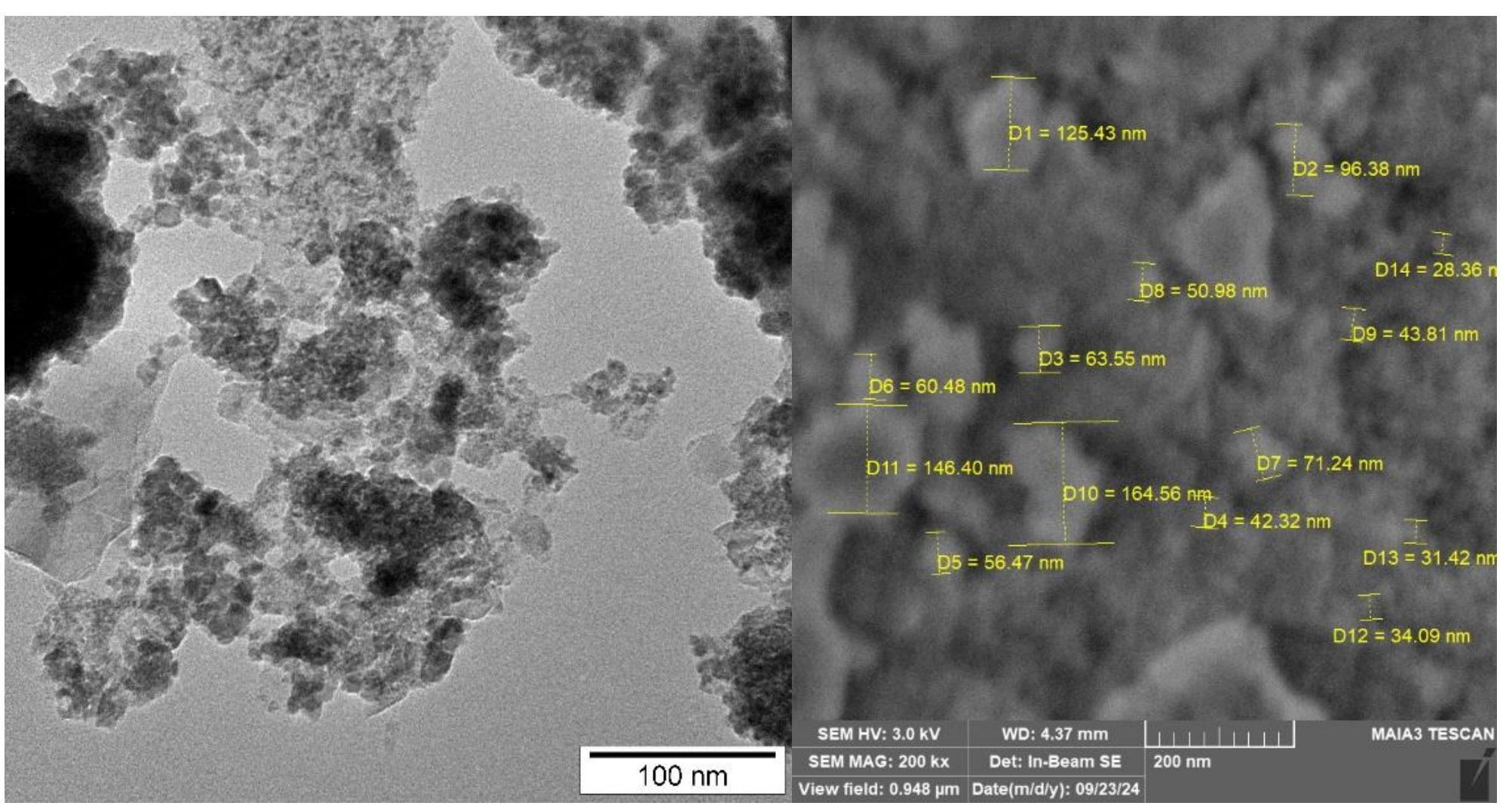

**Figure S1.** TEM image (left) and SEM image (right) of sample N2.

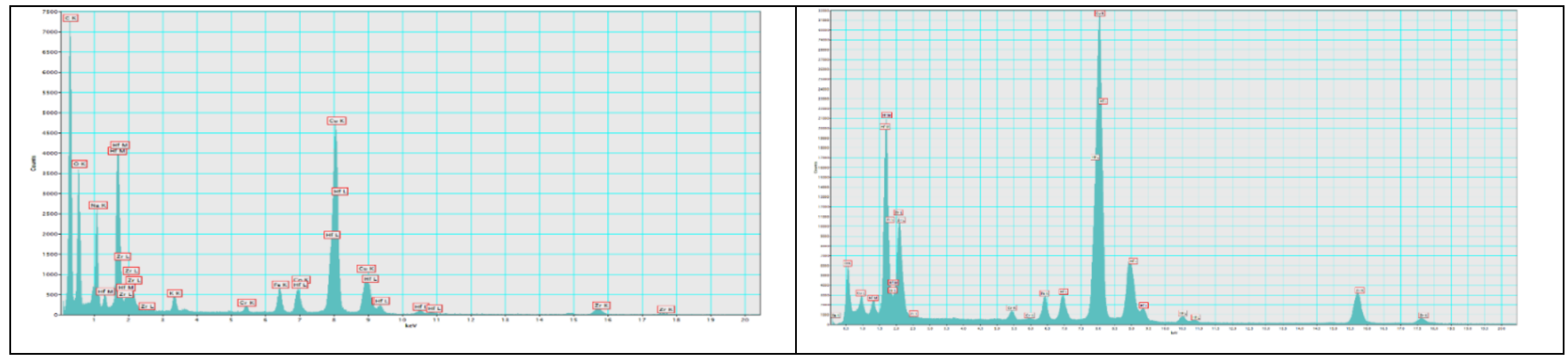

**Figure S2**. EDS spectra of $Hf_{0.5}Zr_{0.5}O_2$ samples annealed in air (left) and in $CO+CO_2$ atmospheres (right).

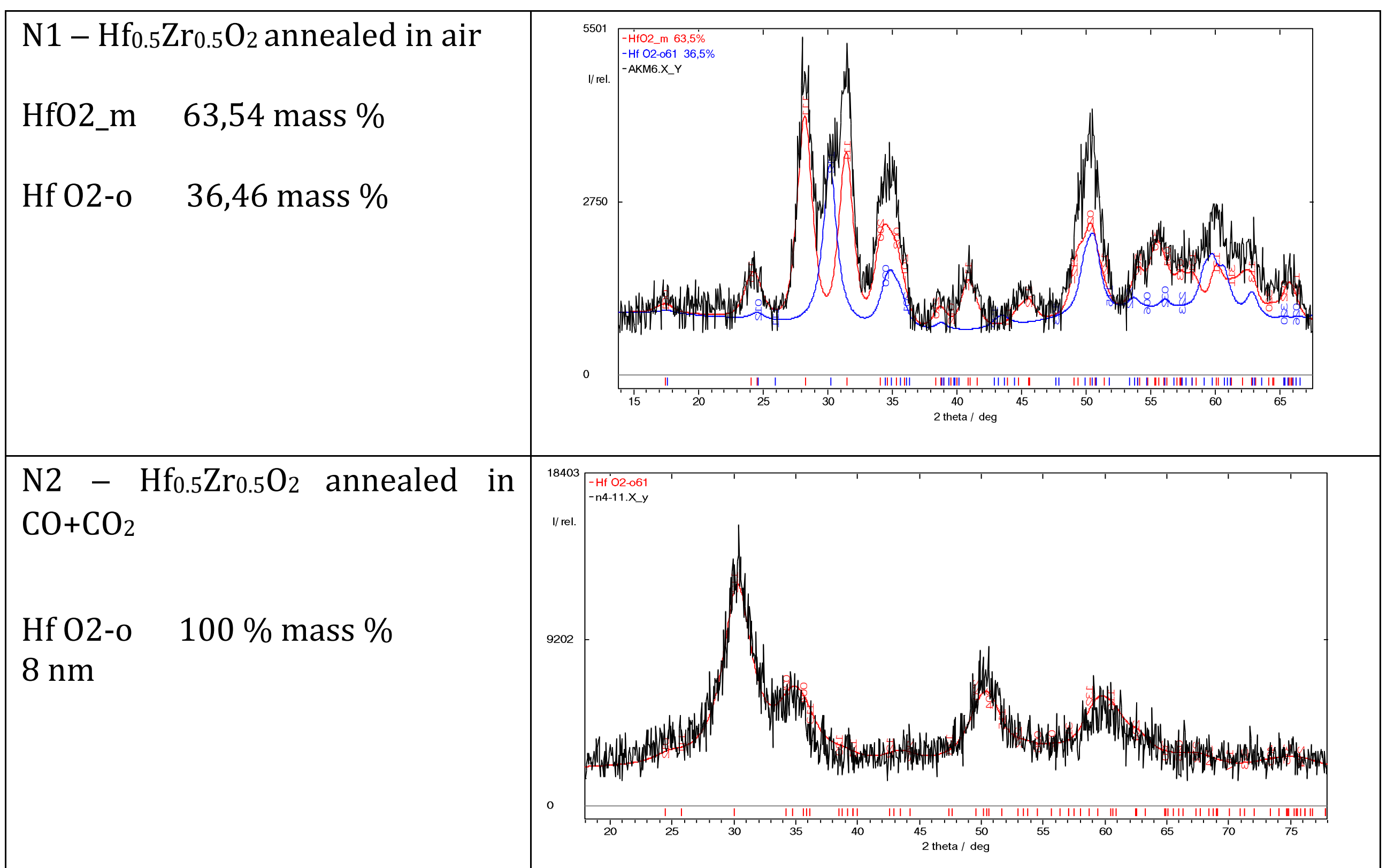

**Figure S3**. XRD spectra of $Hf_{0.5}Zr_{0.5}O_2$ nanoparticles annealed in air **(a)** and $CO+CO_2$ atmospheres **(b)**.

## Supplement S2. Dielectric Measurements of Nanocomposite "(Hf,Zr)$O_2$ Nanoparticles – PVDF Matrix"

The $Hf_{0.5}Zr_{0.5}O_2$ powder occupied approximately 13 vol. % of the total composite volume. The concentration of nanoparticles in PVDF was about 38.5 wt. % in the $Hf_{0.5}Zr_{0.5}O_2$ samples annealed in $CO+CO_2$, and about 39.4 wt. % in the $Hf_{0.5}Zr_{0.5}O_2$ samples annealed in air. The effective dielectric permittivity of the $Hf_{0.5}Zr_{0.5}O_2$ – PVDF nanocomposites is shown in **Fig. S4.**

In the Maxwell–Wagner case, dielectric constants often reach values on the order of tens of thousands at low frequencies, which was not observed in our experiment. At the same time, the relatively small dielectric permittivity ε of PVDF (≈12) complicates the determination of the true permittivity of the composite materials. The effective dielectric permittivity of the composite can be estimated by the Bruggeman's approach, which gives $\varepsilon_{eff}=\frac{1}{4}\left(\beta+\sqrt{\beta^2+8\varepsilon_1\varepsilon_2}\right)$, $\beta=(3\eta_1-1)\varepsilon_1+(3\eta_2-1)\varepsilon_2$,

where $\eta_i$ represents the volume fraction of each constituent and $\varepsilon_i$ corresponds to its dielectric constant. Thus, for a volume fraction of $Hf_{0.5}Zr_{0.5}O_2$ equal to 0.13, the effective dielectric permittivity ranges from about 16 to 19 when $\varepsilon_{HZO}$ varies between 100 and $10^4$. It should be noted that applying the Bruggeman model to experimental data does not yield meaningful dielectric constant values for the studied nanoparticles. This is likely due to the high conductivity of the obtained composites and the tendency of nanoparticles to form irregular clusters, with possible air inclusions between them.

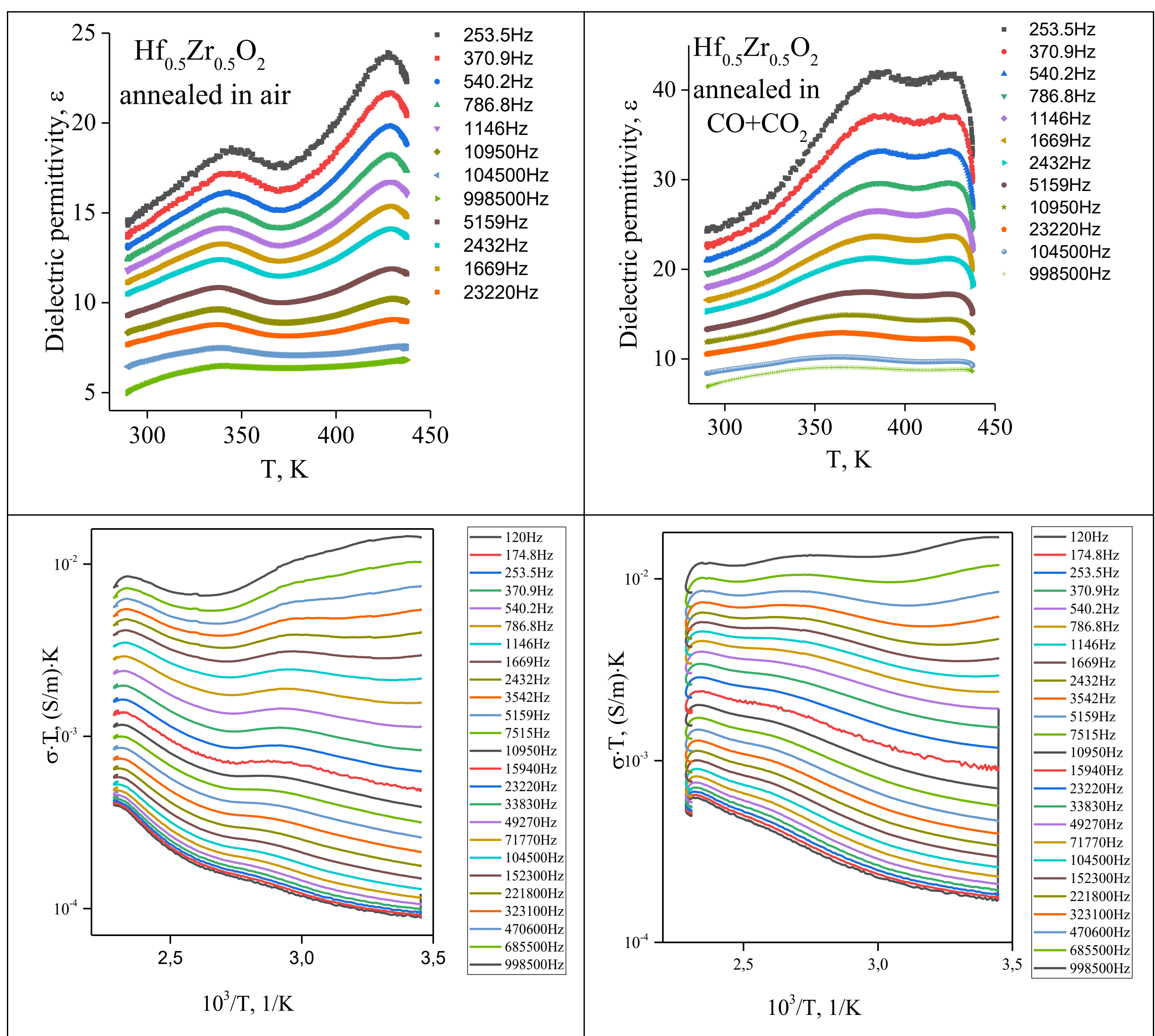

**Figure S4**. Effective dielectric permittivity and conductivity of the $Hf_{0.5}Zr_{0.5}O_2$ – PVDF nanocomposites. The $Hf_{0.5}Zr_{0.5}O_2$ nanopowders, used for the nanocomposite preparation, were annealed in air (left) and in $CO+CO_2$ atmosphere (right).